\documentclass[twocolumn,english,aps,showpacs,prl,superscriptaddress,floatfix]{revtex4-1}
\usepackage{color}
\usepackage{amsmath}
\usepackage{amssymb}
\usepackage{graphicx}
\usepackage{esint}

\newcommand{\w}{\omega}
\newcommand{\mimp}{m_{\rm imp}}
\newcommand{\llangle}{\langle\langle}
\newcommand{\rrangle}{\rangle\rangle}

\graphicspath{{./}{./fig/}}

\begin{document}
\title{
Destruction of long-range order in noncollinear two-dimensional \\ 
antiferromagnets by random-bond disorder
}

\author{Santanu Dey}
\affiliation{Institut f\"ur Theoretische Physik and W\"urzburg-Dresden Cluster of Excellence ct.qmat, Technische Universit\"at Dresden,
01062 Dresden, Germany}
\author{Eric C. Andrade}
\affiliation{Instituto de F\'{i}sica de S\~ao Carlos, Universidade de S\~ao Paulo, C.P. 369,
S\~ao Carlos, SP,  13560-970, Brazil}
\author{Matthias Vojta}
\affiliation{Institut f\"ur Theoretische Physik and W\"urzburg-Dresden Cluster of Excellence ct.qmat, Technische Universit\"at Dresden,
01062 Dresden, Germany}

\begin{abstract}
We consider frustrated Heisenberg antiferromagnets, whose clean-limit ground state is characterized by non-collinear long-range order with non-zero vector chirality, and study the effects of quenched bond disorder, i.e., random exchange couplings. A single bond defect is known to induce a dipolar texture in the spin background independent of microscopic details. Using general analytical arguments as well as large-scale simulations for the classical triangular-lattice Heisenberg model, we show that any finite concentration of such defects destroys long-range order for spatial dimension $d\leq 2$, in favor of a glassy state whose correlation length in $d=2$ is exponentially large for small randomness.
Our results are relevant for a wide range of layered frustrated magnets.
\end{abstract}

\date{\today}
\maketitle


Spatially inhomogeneous exchange couplings are ubiquitous to magnetic solids. Such disorder, usually dubbed random-bond disorder, arises from crystalline defects or intentional chemical substitution on non-magnetic sites, causing local changes in bond lengths or bond angles which in turn influence local exchange couplings.

The effect of bond disorder in magnets has been studied extensively, both experimentally and theoretically, with particular focus on frustrated systems \cite{ramirez94,starykh15} where the delicate balance of partially satisfied constraints can be easily broken by disorder.
In general, systems which are gapped in the clean limit are expected to be stable against weak disorder, such that the phase realized in the clean system survives up to a critical level of disorder where it typically gives way to a spin-glass state \cite{villain79}. The fate of gapless systems is more subtle, and various scenarios are possible: For strongly frustrated systems, weak bond disorder may immediately induce a spin glass, as in the classical pyrochlore Heisenberg antiferromagnet \cite{chalker07,chalker10}, or it may stabilize a distinct disorder-driven long-range-ordered state, as in the classical XY antiferromagnet on the pyrochlore lattice \cite{zhito14} or the frustrated square lattice \cite{henley89}.
In contrast, for weakly frustrated systems it is frequently assumed that the clean-limit order survives the introduction of weak bond disorder, as is the case without frustration.

\begin{figure}[b]
\includegraphics[width=0.95\columnwidth]{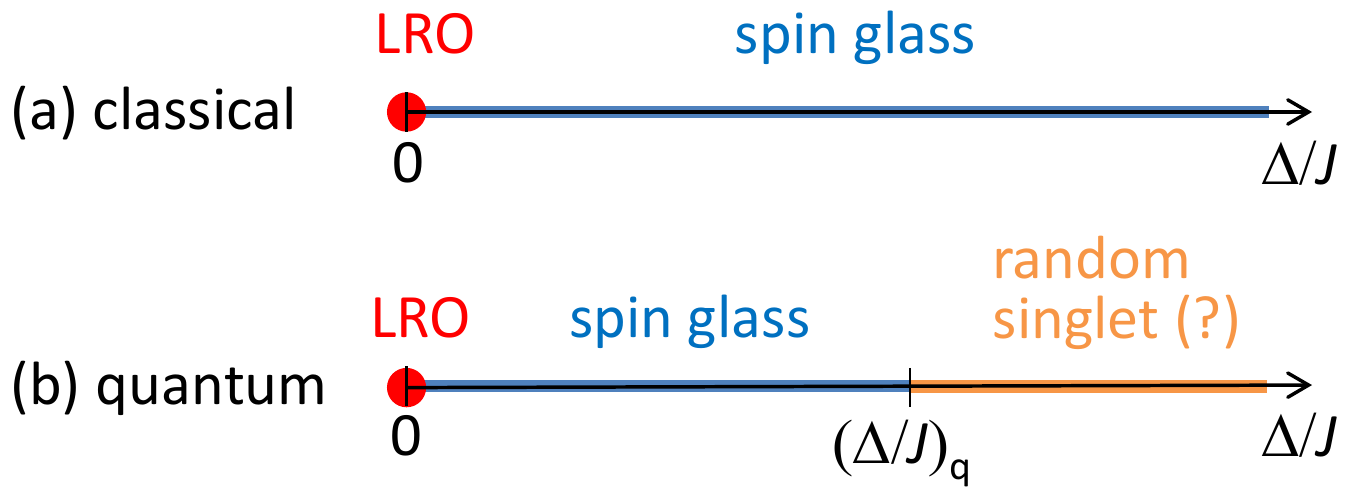}
\caption{
Schematic ground-state phase diagram of chiral non-collinear magnets with quenched bond disorder in space dimension $1<d\leq2$, with $\Delta/J$ parametrizing the disorder strength.
(a) Classical limit ($S\to\infty$): The LRO of the clean system is destroyed for infinitesimal disorder in favor of a spin glass.
(b) Quantum case (conjectured): The spin glass turns into a random-singlet (or valence-bond) state at a critical level of disorder (whose value depends on microscopic details).
}
\label{fig:pd}
\end{figure}

In this Rapid Communication, we argue that long-range order (LRO) is in fact \emph{not} stable against weak bond disorder for an important class of weakly frustrated magnets, namely SU(2)-symmetric non-collinear magnets in two space dimensions, with the triangular-lattice Heisenberg antiferromagnet being a prominent example. A single bond defect induces a dipolar texture in the spin background \cite{syrom15,syrom19}. A finite concentration of defects then corresponds to spatially fluctuating dipoles which we show to destroy ground-state LRO in space dimension $d\leq 2$, Fig.~\ref{fig:pd}(a). The resulting non-coplanar glassy state is characterized by exponentially decaying spin and chirality correlations.
Given that $d=2$ is the lower critical dimension, the magnetic correlation length of this spin glass is exponentially large for weak disorder, implying that both numerical simulations and experiments will detect the destruction of LRO only beyond a resolution-dependent level of bond disorder.
As a byproduct, we show that site dilution has a much weaker effect, leaving bulk LRO intact in the weak-disorder limit, thus invalidating the assertion that bond disorder and site dilution have similar effects.
To connect to experiments, we discuss the physics of finite temperature, weak interlayer coupling, and weak breaking of SU(2) symmetry. In all these cases, the behavior of the clean system survives up to a small finite level of bond randomness.

We note that previous numerical work \cite{kawamura14,kawamura15,sheng18} for the triangular-lattice spin-$1/2$ Heisenberg model suggested that LRO is destroyed only beyond a critical level of bond disorder, in favor of a randomness-dominated spin-liquid-like (or random-singlet) state \cite{dasgupta80,fisher94}. Our results instead imply that true LRO is lost already for infinitesimal bond disorder [Fig.\ref{fig:pd}(b)], but this could not be detected in the numerics because of finite-size effects.


\paragraph{Model and general considerations.---}
To be specific, we will consider the spin-$S$ triangular-lattice Heisenberg model with antiferromagnetic first-neighbor and second-neighbor couplings,
\begin{equation}
\mathcal{H} =
\sum_{\langle ij\rangle} J_{1,ij} \vec{S}_i\cdot\vec{S}_j +
\sum_{\llangle ij\rrangle} J_{2,ij} \vec{S}_i\cdot\vec{S}_j
\end{equation}
whose ground state in the clean limit, $J_{1,ij} \equiv J_1$ and $J_{2,ij} \equiv J_2$, displays coplanar spiral $120^\circ$ LRO with propagation wavevector $\vec{Q}=\pm(4\pi/3,0)$ for $\alpha\equiv J_2/J_1<1/8$. We will vary $\alpha$ to tune the stiffness of the $120^\circ$ LRO.
The $120^\circ$ state is chiral: For spins in the $x$-$y$ plane in spin space, the vector chirality $\vec{S}_i\times\vec{S}_j$ for any given directed pair of sites $i,j$ can point along either $+\hat{z}$ or $-\hat{z}$, corresponding to the two possible propagation wavevectors
\cite{braggfoot}. As a result, inversion symmetry is broken as well, which plays an important role for the defect physics.

We will be interested in the random-bond case where the $J_{1,ij}$ and $J_{2,ij}$ are taken as (independent) random variables. Bond disorder can be made weak either by employing a narrow distribution of $J$, or by having a small concentration of defect bonds deviating from the majority coupling strength.

Our main focus will be on semiclassical spin order --- this is appropriate for the clean system in the presence of LRO, and by continuity also for weakly disordered systems. In this regime, quantum effects will only yield quantitative corrections, and we will show explicit results for the classical case, formally $S\to\infty$.
Strong quantum effects can be expected to be relevant at large disorder and small $S$: Here, the formation of singlet bonds akin to a random-singlet state has been proposed \cite{kawamura14,kawamura15,sheng18}; we will comment on this at the end of this Rapid
Communication.

Most generally, our qualitative results will apply to all two-dimensional Heisenberg magnets with semiclassical non-collinear, but co-planar LRO, both with spontaneous and explicit (i.e. crystallographic) chirality.


\begin{figure}
\begin{centering}
\includegraphics[width=0.43\columnwidth]{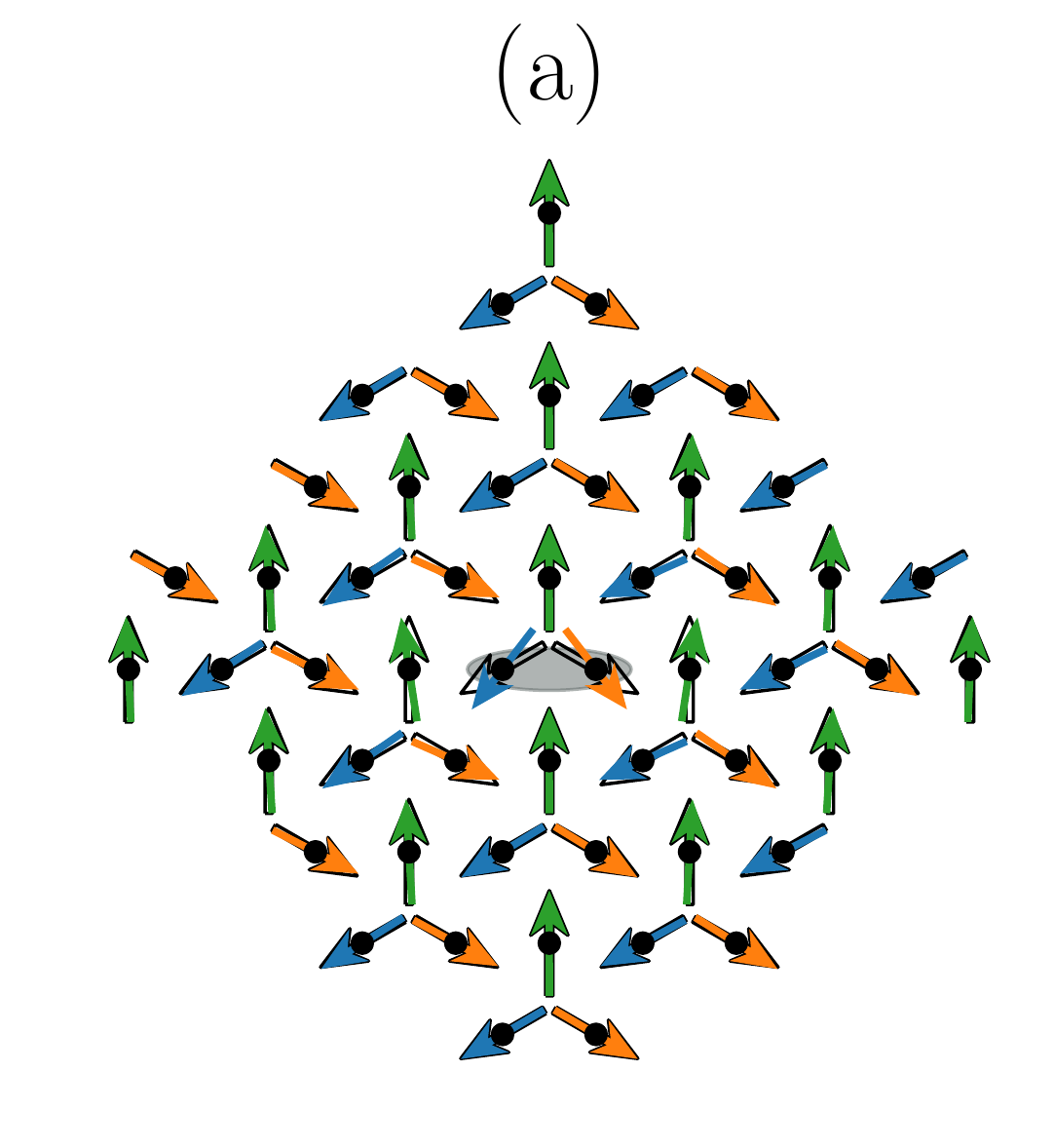}\hfill%
\includegraphics[width=0.52\columnwidth]{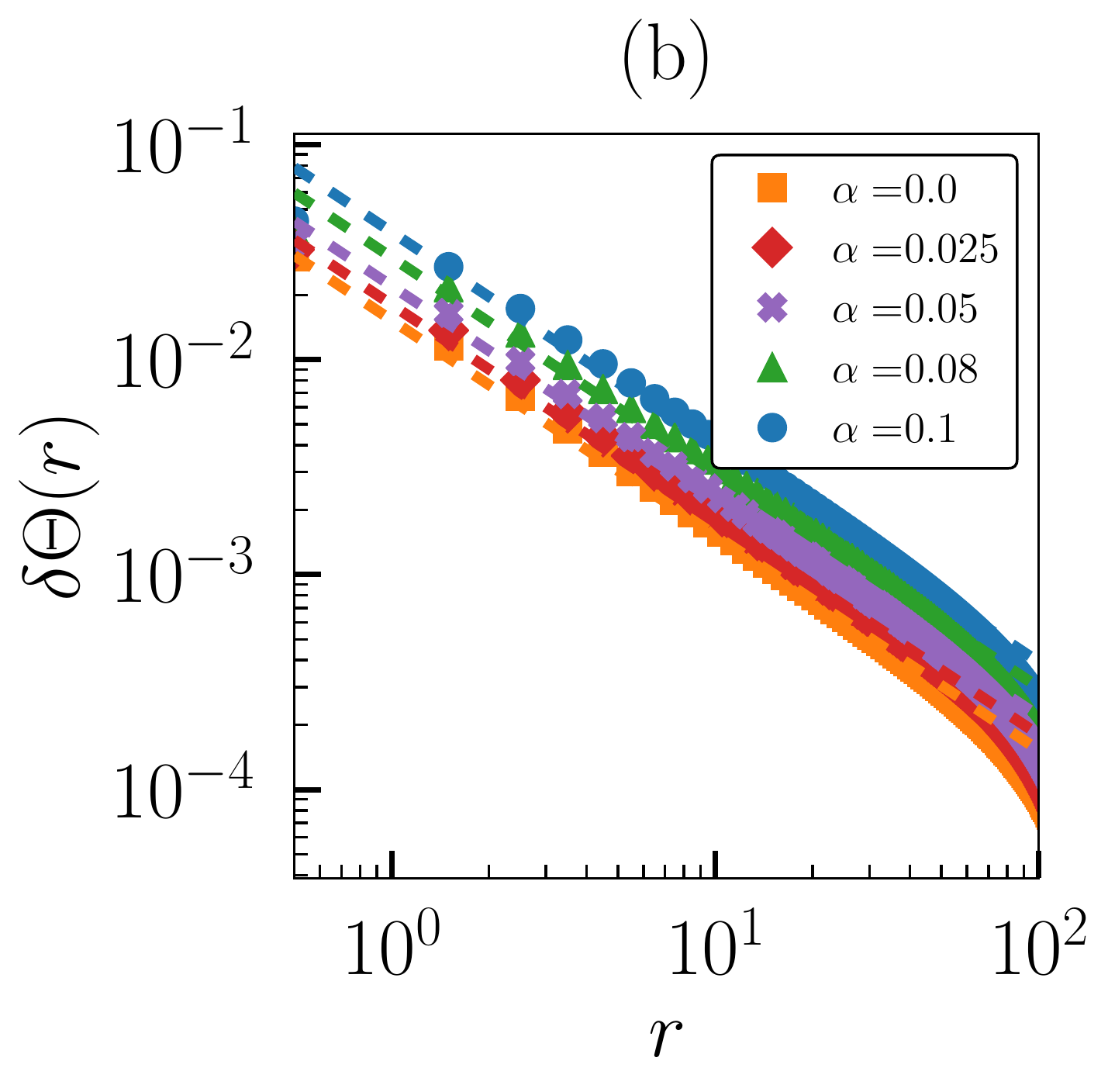}
\end{centering}
\vspace*{-5pt}
\caption{Single bond defect in a triangular-lattice Heisenberg antiferromagnet.
(a) Spin configuration near the defect bond (shaded) for $\delta J=-J_1$ \eqref{onebond} and $J_2=0$; the unperturbed $120^\circ$ order is shown in gray.
(b) Defect-induced spin rotation angles $\delta\Theta$ as a function of distance $r$ from the defect, for $\delta J=-J_1/10$ and different $\alpha\equiv J_2/J_1$. The dashed lines are fits to $1/r$ behavior for $2<r<20$; deviations at large $r$ arise from finite sample size and periodic boundary conditions, $L=300$.
}
\label{fig:single}
\end{figure}

\paragraph{Single bond defect.---}
A single defect bond in an otherwise homogeneous system has been studied before \cite{syrom15,syrom19}, and we summarize the key results. We consider
\begin{equation}
\label{onebond}
J_{1,ij} = \left\{ \begin{array}{ll} J_1 +\delta J & (i,j)=(0,1), \\ J_1 & \rm{otherwise}. \end{array} \right.
\end{equation}
and $J_{2,ij} \equiv J_2$. The presence of the defect locally relieves frustration, leading to a readjustment of the spin directions. Numerical results for a bond defect with $\delta J=-J_1$ are shown in Fig.~\ref{fig:single}(a).
The spin configuration remains coplanar and can be analyzed in terms of angles $\delta\Theta_i$ which describe the defect-induced in-plane rotation of the spins $\vec{S}_i$
with respect to the $120^\circ$ LRO. The Hamiltonian with couplings \eqref{onebond} has inversion symmetry with respect to the center of the defect bond, and $\delta\Theta_i$ is found to be \emph{odd} under this inversion:
$\delta\Theta_i$
has a $p$-wave-like shape and decays proportional to $1/r$, where $r$ is the distance from the defect [Fig.~\ref{fig:single}(b)].
This implies that the bond defect acts as a \emph{dipolar} perturbation; this can be contrasted to the case of a vacancy which induces an octupolar texture, with an $f$-wave shape and a $1/r^3$ decay of $\delta\Theta_i$ \cite{wollny11,wollny12}.

The connection between the local release of frustration and the long-range nature and shape of the distortion can be understood at a linear-response level: The frustration is released such that the two spins $0,1$ on the defect bond align more (less) antiparallel compared to the $120^\circ$ state for $\delta J>0$ ($\delta J<0$), respectively. This rotation can be induced by a locally transverse field, $h^\perp \propto \delta J$, acting on $\vec{S}_0$ and $\vec{S}_1$ with opposite signs. Working in a local frame of spin coordinates such that the order is uniform along the $\hat{x}$ axis, the locally transverse field acts as $h^\perp \sum_{j=0}^1 \beta_j S_j^y$ with $\beta_j=(-1)^j$. The long-distance pattern follows as the response of the ordered state to this \emph{local dipolar} field.
The relevant in-plane susceptibility $\chi^\parallel(\vec{q})$ is dominated by the in-plane spin-wave modes whose dispersion is $\w_q\propto|\vec{q}|$ such that $\chi^\parallel(\vec{q}) = N_0^2/(\rho_s q^2)$ where $\rho_s$ is the spin stiffness against in-plane twists, and $N_0$ is the magnitude of the order parameter \cite{ssbook}, both taken for the clean system.
%
This yields, in the continuum limit, $\delta\Theta(\vec r) \propto \int d^d q e^{i\vec{q}\cdot\vec{r}} \beta_{\vec{q}} \chi^\parallel(\vec{q})$ in $d$ space dimensions where $\beta_{\vec{q}} = \hat{e} \cdot \vec{q}$ is the $p$-wave form factor of the local perturbation, with $\hat{e}$ being the lattice vector of the directed defect bond. Together, we obtain
\begin{equation}
\label{eq:texture}
\delta\Theta(\vec r) = \kappa \, \delta J \, \frac{N_0^2}{\tilde\rho_s} \, \frac{\hat{e} \cdot \vec{r}}{r^d}
\end{equation}
where $\tilde\rho_s=\rho_s/\mathcal{A}$ with $\mathcal{A}$ the unit-cell area \cite{stifffoot}, and $\kappa$ a numerical prefactor, see Ref.~\onlinecite{suppl} for details. Being inversely proportional to the stiffness, the defect response thus depends significantly on $\alpha=J_2/J_1$, consistent with Fig.~\ref{fig:single}(b).

The fact that a bond defect produces an antisymmetric (or inversion-odd) texture is intimately connected to the chirality of the ground state: The Hamiltonian itself is inversion-even, and a single defect cannot spontaneously break this symmetry. However, the ground state is chiral, with broken inversion symmetry, enabling the defect to produce an odd perturbation. In other words, the ground-state chirality endows the bond defect with a direction, as required for a dipole, and reversing the chirality will reverse the sign of the dipole, $\hat{e} \leftrightarrow -\hat{e}$.

As an aside, we note that the state with a single bond defect has a finite uniform magnetization, $\mimp$, which takes a non-universal fractional value, similar to the vacancy case \cite{wollny11}. For the bond defect with $\delta J=-J_1$ and $\alpha=0$ we have found
$\mimp/S = 0.396 + \mathcal{O}(S^{-1})$. We also note that a single (weak) defect on a second-neighbor bond has no effect on the classical spin order, as second-neighbor spins are parallel in the $120^\circ$ state.


\paragraph{Destruction of LRO by dipolar fluctuations.---}
We now turn to the case of a finite defect concentration and argue that LRO is generically destroyed for $d\leq 2$, adopting an argument originally due to Aharony~\cite{aharony78}. To this end, we assume LRO and employ local rotated frames as above, with order along the $\hat{x}$ axis.
We distribute random bonds $J_{1,ij}$ on the lattice and define $\delta J_{ij}=J_{1,ij}-J_1$ where $J_1=\overline{J_{1,ij}}$ is the disorder-averaged $J_1$, such that the disorder strength is parameterized by $\overline{\delta J_{ij}^2} \equiv \Delta^2$.
For each directed bond $\hat{e}_{ij}$, $\vec{d}_{ij} \equiv \hat{e}_{ij}\delta J_{ij}$ takes the role of a local dipole strength.
For weak randomness, the resulting rotation of an individual spin is proportional to its transverse magnetization in the rotated frame, $\delta\Theta_l = \langle S_{l}^{\perp} \rangle / N_0$. It can be estimated using the linear response as above
\begin{align}
\left\langle S_{l}^{\perp}\right\rangle = \kappa \frac{N_0^3}{\tilde\rho_s}
\sum_{\langle ij\rangle}
\frac{\vec{d}_{ij}\cdot \vec{r}_{l,ij}}{r_{l,ij}^{d}}
\end{align}
where $\vec{r}_{l,ij}$ is the vector connecting site $l$ and the center of the $ij$ bond.
The disorder-averaged transverse magnetization, $\overline{\langle S_{j}^{\perp}\rangle}$, vanishes because the averaged dipole strength has zero mean. In contrast, the averaged magnetization correlation function is non-zero \cite{suppl},
\begin{equation}
\overline{\left\langle {S_{l}^{\perp}}^2\right\rangle } = \tilde\kappa \, \Delta^2 \frac{N_0^6}{\tilde\rho_s^2} \int dr \,r^{1-d}
\label{fluctint}
\end{equation}
where we have passed to the continuum limit, the angular average has been performed,
and $\tilde\kappa \propto \kappa^2$ is a prefactor.

The above integral is infrared divergent for $d\leq 2$, such that the local transverse magnetization fluctuations diverge for $d\leq2$: These fluctuations, arising from dipolar bond disorder and transmitted by long-wavelength modes, then destroy the assumed ordered state. In contrast, for $d>2$ a finite defect concentration is required to destroy order.
The same conclusion can be reached by a more elaborate renormalization-group (RG) treatment of a relevant non-linear sigma model (for details see Ref.~\onlinecite{suppl}). The destruction of LRO by dipolar fluctuations has been studied in different contexts before \cite{cherepanov99,hasselmann04}.


\paragraph{Emergent spin glass.---}
We now discuss the nature of the emerging zero-temperature state without magnetic LRO. In the semiclassical limit, this must be a state with spontaneously broken SU(2) symmetry and, given the random-field character of the problem, is a spin glass with frozen short-ranged spin order \cite{fischer,xy_foot}.

\begin{figure}[t]
\includegraphics[width=0.5\columnwidth]{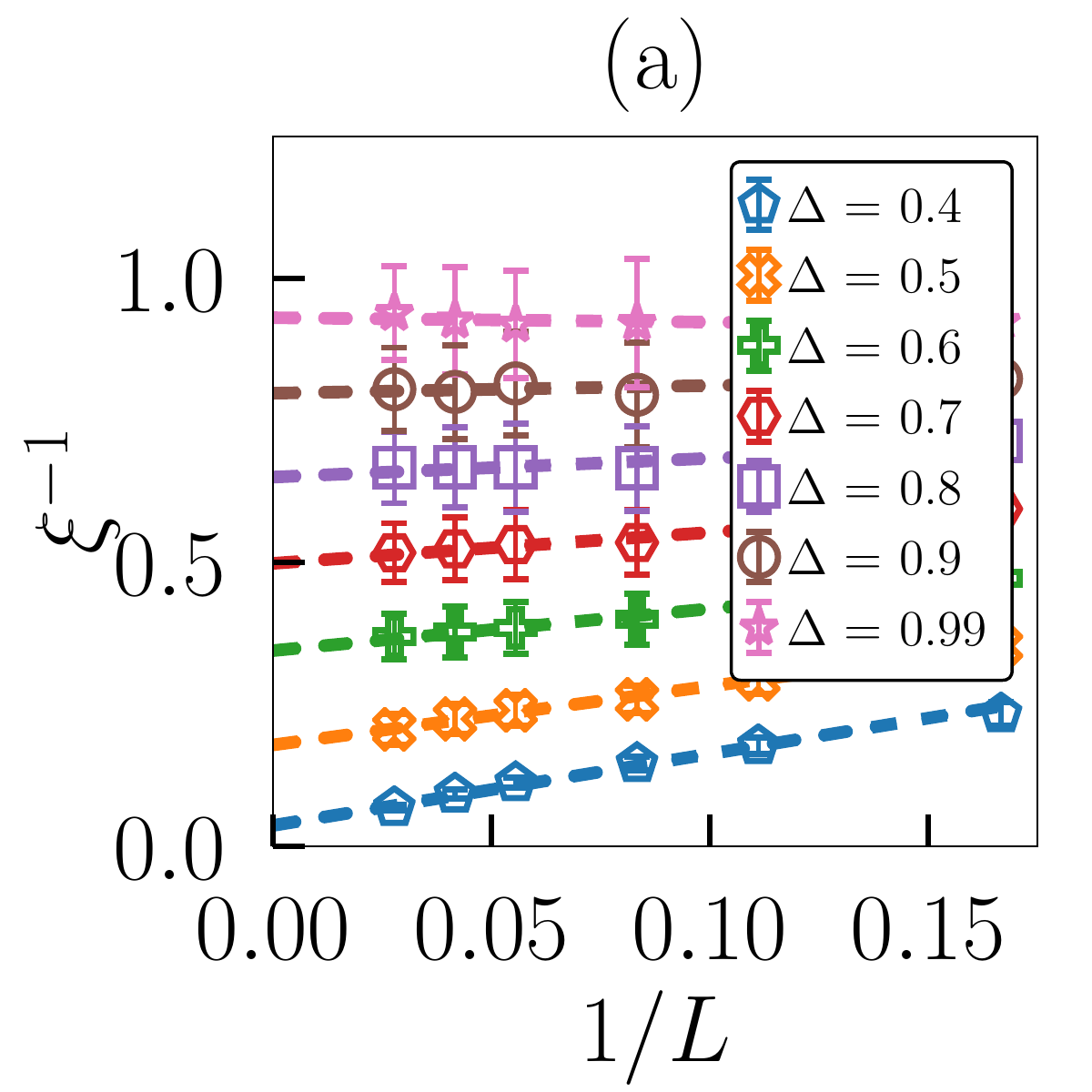}%
\includegraphics[width=0.5\columnwidth]{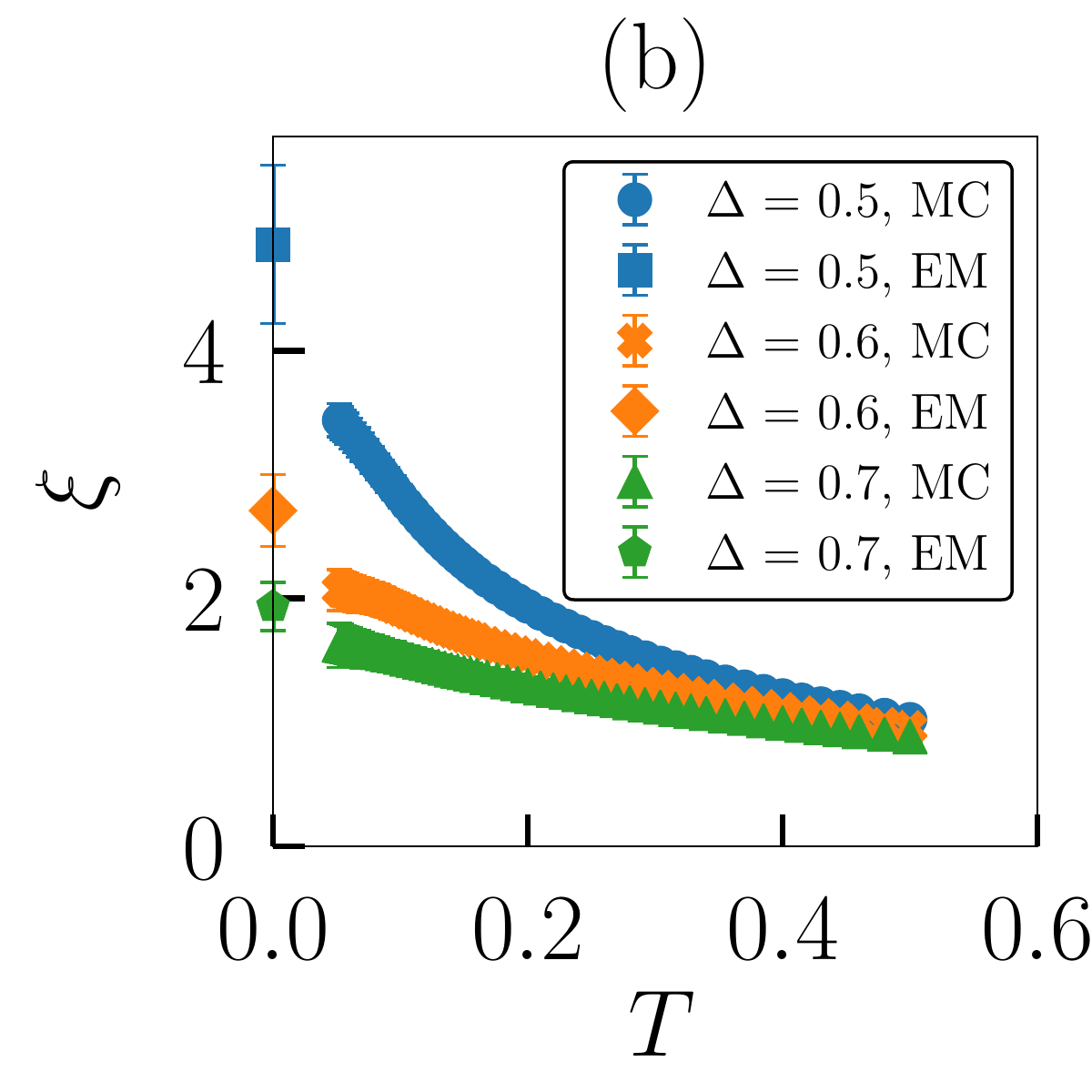}
\includegraphics[width=0.5\columnwidth]{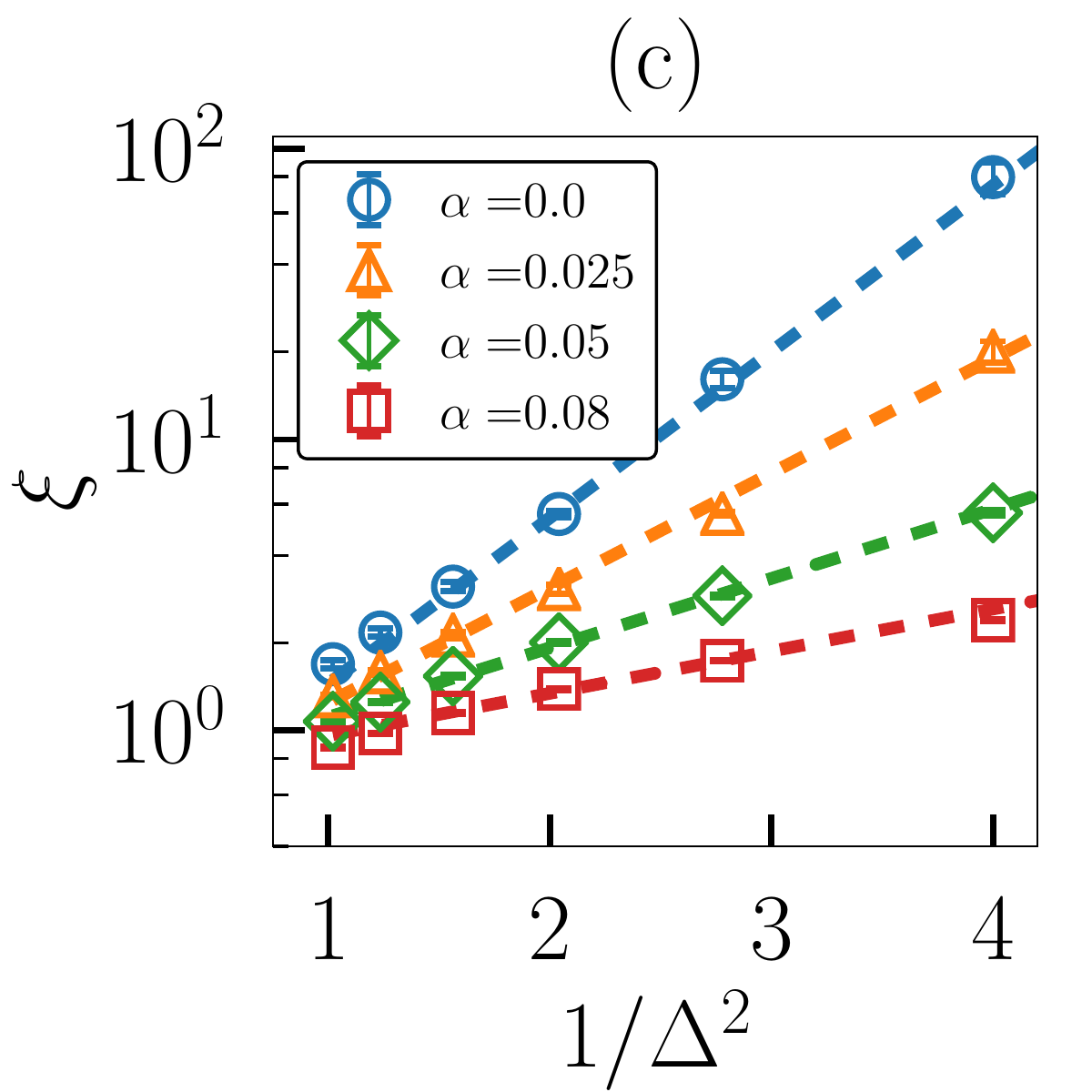}%
\includegraphics[width=0.5\columnwidth]{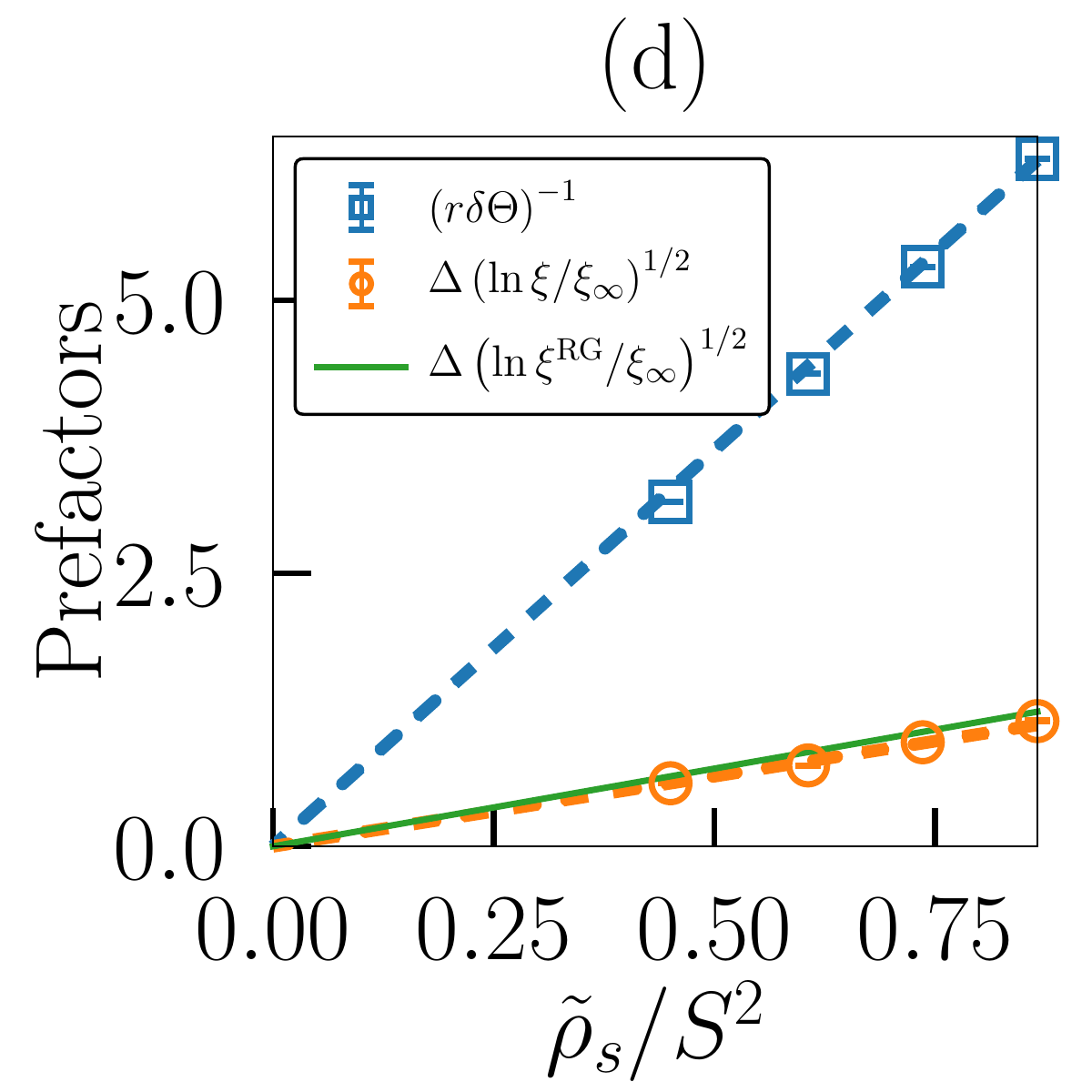}
\caption{
Magnetic correlation length $\xi$ for the triangular-lattice Heisenberg model with Gaussian bond disorder.
(a) Finite-size scaling of $1/\xi$ as function of linear system size $L$ for $\alpha=0.05$, $T=0$, and different disorder strength $\Delta$.
(b) Comparison of $\xi$ obtained from finite-$T$ MC simulation and $T=0$ energy minimization (EM), for $\alpha=0.05$ and $L=36$.
(c) $\ln\xi$ plotted as function of $1/\Delta^2$, illustrating the scaling from Eq.~\eqref{xi}.
    (d) Test of stiffness dependence: $[\ln(\xi/\xi_\infty)]^{1/2}\Delta$ and single-defect $1/[r\delta\Theta(r)]$ for different $\alpha$, plotted as function of spin stiffness. The solid line corresponds to the RG result for $\xi$.
Dashed lines are linear fits.
}
\label{fig:xi}
\end{figure}

We can estimate its magnetic correlation length $\xi$ simply by assuming that $\xi$ provides an infrared cut-off to the integral Eq.~\eqref{fluctint}, i.e., by identifying $\xi$ with a domain size. The stability condition $\overline{\langle {S_{l}^{\perp}}^2\rangle } \lesssim N_0^2$ translates into $\xi^{2-d} \propto \tilde\rho_s^2/(\Delta^2 N_0^4)$ for $d<2$, whereas in $d=2$ the correlation length is exponentially large for small $\Delta$,
\begin{equation}
\ln \frac{\xi}{\xi_\infty} \propto \frac{\tilde\rho_s^2}{\Delta^2 N_0^4}
\label{xi}
\end{equation}
where the constant $\xi_\infty$ formally represents the correlation length as $\Delta\to\infty$.
While the spin configurations remain coplanar in the limit of small $\Delta$, this is no longer true at finite disorder \cite{suppl}.


\paragraph{Numerical results for finite disorder.---}
Our analytical results are well borne out by large-scale simulations for the bond-disordered triangular-lattice $J_1$-$J_2$ Heisenberg model, using both ground-state energy minimization and finite-temperature Monte Carlo (MC) techniques \cite{suppl}.
We have employed different disorder distributions; below we show results where the $J_{1,ij}$ and $J_{2,ij}$ have been drawn independently from Gaussian distributions with mean $J_1$ (which we use as unit of energy) and $J_2=\alpha J_1$ and widths $\Delta$ and $\alpha\Delta$, respectively.

We have determined the magnetic correlation length $\xi$ from the disorder-averaged static structure factor. An example for the finite-size scaling of $\xi$ at $T=0$ is shown in Fig.~\ref{fig:xi}(a). For large disorder, we clearly detect the absence of LRO, while for small disorder the correlation length exceeds the available system sizes, such that the finite-size scaling is inconclusive. The temperature dependence of $\xi$ obtained via MC simulations is consistent with the data at $T=0$, Fig.~\ref{fig:xi}(b). The finite-$T$ spin-glass correlation length is found much larger than its magnetic counterpart, consistent with a spin-glass ground state. An analysis of individual spin configurations confirms that the glassy state is non-coplanar \cite{suppl}.

Plotting the extrapolated data for $\xi$ at $T=0$ as function of disorder strength $\Delta$, Fig.~\ref{fig:xi}(c), we find the $\Delta$ dependence perfectly consistent with the exponential behavior predicted by Eq.~\eqref{xi}. This strongly suggests that LRO is indeed destroyed for \textit{any} non-zero $\Delta$.
Equation ~\eqref{xi} also predicts that $[\ln(\xi/\xi_\infty)]^{1/2}\Delta$ is proportional to the clean-limit stiffness which reads $\tilde{\rho}_s = S^2 (J_1 - 6 J_2) \sqrt{3}/2$ in the classical limit \cite{suppl}. This proportionality is tested in Fig.~\ref{fig:xi}(d) and found to be perfectly obeyed. In fact, the RG treatment \cite{suppl} also yields an approximate expression of the proportionality factor which agrees with the data. In addition, Fig.~\ref{fig:xi}(d) also shows $1/[r \delta\Theta(r)]$ for the single-defect texture which is proportional to the stiffness, in quantitative agreement with Eq.~\eqref{eq:texture}.
Finally, we do not detect any crossing points in the $\Delta$ dependence of both the magnetic correlation length and the Binder parameter for different $L$ \cite{suppl}. This confirms the absence of a critical $\Delta$ and thus the phase diagram in Fig.~\ref{fig:pd}(a).


\paragraph{Perturbations: Finite $T$, interlayer coupling, and anisotropies.---}
Beyond the two-dimensional Heisenberg model discussed so far, a number of effects are important.
First, in a strictly two-dimensional SU(2)-symmetric system, the clean-limit LRO is restricted to $T\!=\!0$ due to the Mermin-Wagner theorem. Hence, the clean system is short-range-ordered at any finite $T$, with the thermal correlation length $\xi_T$ scaling as $\ln \xi_T \propto \rho_s/T$ \cite{chakravarty89}. Now, bond disorder limits the correlation length according to Eq.~\eqref{xi}, which defines a disorder-dependent crossover temperature which scales quadratically with the disorder level, $T^\ast \propto \Delta^2 N_0^4/\rho_s$, below which the system settles into its $T\!=\!0$ glassy state.
We note that $d=2$ is below the lower critical dimension for spin-glass order \cite{parisi17}, hence, there will be no thermodynamic glass transition in a strictly two-dimensional system.
Second, a small but finite interlayer coupling will render the system three-dimensional at sufficiently low temperature, leading to finite-$T$ LRO which is also stable against weak bond disorder. As a result, LRO is destroyed in favor of a glassy state only beyond a critical level of disorder which scales with the interlayer coupling according to $J_\perp/J \propto \Delta^4 N_0^8/\rho_s^4$.
Third, if SU(2) is broken down at the Hamiltonian level such that there is no spin rotation symmetry in the ordering plane, the in-plane mode of the clean system acquires a gap. As a result, the texture induced by a single defect decays exponentially. The system displays again LRO at low $T$ which is stable against weak bond disorder. Here, the critical level of disorder scales logarithmically with the gap (for details see Ref.~\onlinecite{suppl}).


\paragraph{Quantum effects.---}
So far our analysis was based on semiclassical spin order. It is strictly valid for $S\to\infty$, but qualitatively also applies to finite $S$: $1/S$ corrections to observables can be calculated and are generically non-singular at $T\!=\!0$ \cite{wollny11,suppl}. However, for small $S$ the physics can change; in particular, local spin order can be destroyed by quantum fluctuations.

For the non-collinear magnet at hand, it is conceivable that a finite amount of bond disorder can lead to the suppression of local order via the emergence of a disorder-dominated valence-bond state \cite{kimchi18,sandvik18}, similar to the celebrated random-singlet state in one dimension \cite{dasgupta80,fisher94}. In fact, the emergence of such a state has been proposed on the basis of numerical simulations for the bond-disordered Heisenberg model both on the triangular and honeycomb lattices \cite{kawamura14,kawamura15,sheng18}.
Together with our insight that infinitesimal bond disorder destroys non-collinear LRO in favor of a spin glass, we conjecture the phase diagram in Fig.~\ref{fig:pd}(b), where the glass gives way to a random-singlet state at large disorder.


\paragraph{Conclusions.---}
Combining analytical arguments and large-scale simulations, we have shown that random-bond defects destroy long-range magnetic order in the ground state of two-dimensional non-collinear antiferromagnets with SU(2) spin symmetry. The key insight, demonstrated explicitly for the triangular-lattice Heisenberg model, is that a finite concentration of effectively dipolar defects destabilize LRO even for weak disorder in favor of a spin-glass state.

Remarkably, the effect of random site dilution in the same system is much weaker: Vacancies induce an octupolar texture \cite{wollny11,wollny12}, and LRO is stable against a small vacancy concentration because the integral corresponding to Eq.~\eqref{fluctint} is {\em not} divergent in this case \cite{maryasin13}.

Our analysis shows that none of the two cases can be described as random-mass disorder in the relevant low-energy field theory for the ordered state, because this would miss the defect-induced random transverse fields. This underlines a fundamental difference between the present non-collinear magnets and their unfrustrated collinear counterparts, where both random-bond disorder and site dilution correspond to random-mass terms in the field theory. We note that bond disorder also tends to destroy LRO in frustrated magnets with strong spin-orbit coupling where defect-induced random transverse fields are even stronger \cite{chernyshev17,hoyos18}.

We expect our ideas to motivate further studies into different types of quenched disorder in weakly frustrated magnets. Investigations for non-coplanar order are underway. On the experimental front, our results are pertinent to many materials such as
Ba$_3$CoSb$_2$O$_9$ \cite{shirata12},
Ba$_3$NiSb$_2$O$_9$ \cite{doi04},
Ba$_8$CoNb$_6$O$_{24}$ \cite{cui18},
Cs$_2$CuCl$_4$ \cite{coldea03},
$\alpha$-CaCr$_2$O$_4$ \cite{toth11}, and
NaCrO$_2$ \cite{hsieh08}.
They also apply to certain iron pnictides where a non-collinear vortex crystal state has been detected \cite{canfield18}.


\begin{acknowledgments}
We are grateful to W. Brenig, J. A. Hoyos, S. Rachel, and S. Trebst for instructive discussions and collaborations on related work.
S.D. and M.V. acknowledge financial support from the Deutsche Forschungsgemeinschaft through SFB 1143 (project-id 247310070) and the W\"urzburg-Dresden Cluster of Excellence on Complexity and Topology in Quantum Matter -- \textit{ct.qmat} (EXC 2147, project-id 390858490). E.C.A. was supported by CNPq (BR) Grant No. 302065/2016-4.
\end{acknowledgments}



\vspace{-5pt}
\begin{thebibliography}{99}
\vspace{-5pt}

\bibitem{ramirez94}
A. P. Ramirez,
Ann. Rev. Mat. Sci. {\bf 24}, 453 (1994).

\bibitem{starykh15}
O. A. Starykh,
Rep. Prog. Phys. {\bf 78}, 052502 (2015).

\bibitem{villain79}
J. Villain,
Z. Phys. B \textbf{33}, 31 (1979).

\bibitem{chalker07}
T. E. Saunders and J. T. Chalker,
Phys. Rev. Lett. \textbf{98}, 157201 (2007).

\bibitem{chalker10}
A. Andreanov, J. T. Chalker, T. E. Saunders, and D. Sherrington,
Phys. Rev. B \textbf{81}, 014406 (2010).

\bibitem{zhito14}
V. S. Maryasin and M. E. Zhitomirsky,
Phys. Rev. B \textbf{90}, 094412 (2014).

\bibitem{henley89}
C. L. Henley,
Phys. Rev. Lett. \textbf{62}, 2056 (1989).

\bibitem{syrom15}
O. I. Utesov, A. V. Sizanov, and A. V. Syromyatnikov,
Phys. Rev. B \textbf{92}, 125110 (2015).

\bibitem{syrom19}
O. I. Utesov and A. V. Syromyatnikov,
arXiv:1902:05820.

\bibitem{kawamura14}
K. Watanabe, H. Kawamura, H. Nakano, and T. Sakai,
J. Phys. Soc. Jpn. \textbf{83}, 034714 (2014).

\bibitem{kawamura15}
T. Shimokawa, K. Watanabe, and H. Kawamura,
Phys. Rev. B \textbf{92}, 134407 (2015).

\bibitem{sheng18}
H.-Q. Wu, S.-S. Gong, and D. N. Sheng,
Phys. Rev. B \textbf{99}, 085141 (2019).

\bibitem{dasgupta80}
C. Dasgupta and S.-k. Ma,
Phys. Rev. B \textbf{22}, 1305 (1980).

\bibitem{fisher94}
D. S. Fisher,
Phys. Rev. B \textbf{50}, 3799 (1994).


\bibitem{braggfoot}
The static spin structure factor of the 120$^\circ$ LRO displays Bragg peaks at both $\pm(4\pi/3,0)$, i.e., is insensitive to chirality.

\bibitem{wollny11}
A. Wollny, L. Fritz, and M. Vojta,
Phys. Rev. Lett. {\bf 107}, 137204 (2011).

\bibitem{wollny12}
A. Wollny, E. C. Andrade, and M. Voj\-ta,
Phys. Rev. Lett. {\bf 109}, 177203 (2012).

\bibitem{ssbook}
S. Sachdev,
{\it Quantum Phase Transitions} (2nd ed.),
Cambridge University Press, Cambridge (2010).

\bibitem{stifffoot}
In the literature, different prefactor conventions are used for the spin stiffness, depending on whether the underlying calculations is lattice-based or continuum-based \cite{suppl}.

\bibitem{suppl}
See supplementary material, which also contains Refs.~\onlinecite{dombre89,chubukov94b,lecheminant95,deutscher93,azaria92,azaria93,fisher85,polyakov-book,glazman90,walker77,alonso96,partemp}, for details on the linear-response calculations, a sketch of the renormalization-group treatment, and further numerical results.

\bibitem{aharony78}
A. Aharony,
Solid State Commun. \textbf{28}, 667 (1978).

\bibitem{cherepanov99}
V. Cherepanov, I. Y. Korenblit, A. Aharony, and O. Entin-Wohlman,
Eur. Phys. J. B \textbf{8}, 511 (1999).

\bibitem{hasselmann04}
N. Hasselmann, A. H. Castro Neto, and C. Morais Smith,
Phys. Rev. B \textbf{69}, 014424 (2004).

\bibitem{fischer}
K. H. Fischer and J. A. Hertz, \emph{Spin Glasses}
(Cambridge University Press, Cambridge, 1991).

\bibitem{xy_foot}
The arguments for the destruction of LRO apply identically to the bond-disordered triangular-lattice XY (instead of Heisenberg) antiferromagnet; however, the resulting state at small disorder displays quasi-long-range order [S. Dey \textit{et al.}, unpublished.]



\bibitem{chakravarty89}
S. Chakravarty, B. I. Halperin, and D. R. Nelson
Phys. Rev. B \textbf{39}, 2344 (1989).

\bibitem{parisi17}
A. Maiorano and G. Parisi,
preprint arXiv:1711.05590.

\bibitem{kimchi18}
I. Kimchi, A. Nahum, and T. Senthil,
Phys. Rev. X \textbf{8}, 031028 (2018).

\bibitem{sandvik18}
L. Liu, H. Shao, Y.-C. Lin, W. Guo, and A. W. Sandvik,
Phys. Rev. X \textbf{8}, 041040 (2018).

\bibitem{maryasin13}
Magnetic LRO in the triangular-lattice Heisenberg model with vacancies is consistent with the results reported in:
V. S. Maryasin and M. E. Zhitomirsky,
Phys. Rev. Lett. \textbf{111}, 247201 (2013).

\bibitem{chernyshev17}
Z. Zhu, P.  A. Maksimov, S. R. White, and A.  L. Chernyshev,
Phys. Rev. Lett. \textbf{119}, 157201 (2017).

\bibitem{hoyos18}
E. C. Andrade, J. A. Hoyos, S. Rachel, and M. Voj\-ta,
Phys. Rev. Lett. {\bf 120}, 097204 (2018).

\bibitem{shirata12}
Y. Shirata, H. Tanaka, A. Matsuo, and K. Kindo,
Phys. Rev. Lett. \textbf{108}, 057205 (2012).

\bibitem{doi04}
Y. Doi, Y. Hinatsu, and K. Ohoyama,
J. Phys.: Condens. Matter \textbf{16}, 8923 (2004).

\bibitem{cui18}
Y. Cui, J. Dai, P. Zhou, P. S. Wang, T. R. Li, W. H. Song, J. C. Wang, L. Ma, Z. Zhang, S. Y. Li, G. M. Luke, B. Normand, T. Xiang, and W. Yu,
Phys. Rev. Materials \textbf{2}, 044403 (2018).

\bibitem{coldea03}
R. Coldea, D. A. Tennant, and Z. Tylczynski,
Phys. Rev. B \textbf{68}, 134424 (2003).

\bibitem{toth11}
S. Toth, B. Lake, S. A. J. Kimber, O. Pieper, M. Reehuis, A. T. M. N. Islam, O. Zaharko, C. Ritter, A. H. Hill, H. Ryll, K. Kiefer, D. N. Argyriou, and A. J. Williams,
Phys. Rev. B \textbf{84}, 054452 (2011).

\bibitem{hsieh08}
D. Hsieh, D. Qian, R. F. Berger, D. J. Cava, J. W. Lynn, Q. Huang, and M. Z. Hasan,
Physica B \textbf{403}, 1341 (2008).

\bibitem{canfield18}
W. R. Meier \textit{et al.},
npj Quant. Mater. \textbf{3}, 5 (2018).



\bibitem{dombre89}
T. Dombre and N. Read,
Phys. Rev. B \textbf{39}, 6797 (1989).


\bibitem{chubukov94b}
A. V. Chubukov, S. Sachdev, and T. Senthil,
J. Phys.: Condens. Matter \textbf{6}, 8891 (1994).

\bibitem{lecheminant95}
P. Lecheminant, B. Bernu, C. Lhuillier, and L. Pierre,
Phys. Rev. B \textbf{52}, 9162 (1995).

\bibitem{deutscher93}
R. Deutscher and H. U. Everts,
Z. Phys. B \textbf{93}, 77 (1993).

\bibitem{azaria92}
P. Azaria, B. Delamotte, and D. Mouhanna,
Phys. Rev. Lett. \textbf{68}, 1762 (1992).

\bibitem{azaria93}
P. Azaria, B. Delamotte, F. Delduc, and T. Jolicoeur,
Nucl. Phys. B \textbf{408}, 485 (1993).

\bibitem{fisher85}
        D. S. Fisher,
        Phys. Rev. B \textbf{31}, 7233 (1985).


    \bibitem{polyakov-book}
        A. M. Polyakov, Gauge Fields and Strings,
        Harwood Publishers (1991)

\bibitem{glazman90}
L. I. Glazman and A. S. Ioselevich,
Z. Phys. B \textbf{80}, 133 (1990).

    \bibitem{walker77}
        L. R. Walker and R. E. Walstedt,
        Phys. Rev. Lett. \textbf{38}, 514 (1977).

    \bibitem{alonso96}
    J. L. Alonso, A. Tarancon, H. G. Ballesteros, L. A. Fernandez, V. Martin-Mayor, and A. Munoz Sudupe,
    Phys. Rev. B \textbf{53}, 2537 (1996).

    \bibitem{partemp}
    K. Hukushima and K. Nemoto,
    J. Phys. Soc. Jpn. \textbf{65}, 1604 (1996).


\end{thebibliography}
\end{document}


\title{Supplementary information for:\\
Destruction of long-range order in non-collinear two-dimensional antiferromagnets by
random-bond disorder
}

\author{Santanu Dey}
\affiliation{Institut f\"ur Theoretische Physik and W\"urzburg-Dresden Cluster of Excellence ct.qmat, Technische Universit\"at Dresden, 01062 Dresden, Germany}
\author{Eric C. Andrade}
\affiliation{Instituto de F\'{i}sica de S\~ao Carlos, Universidade de S\~ao Paulo, C.P. 369, S\~ao Carlos, SP,  13560-970, Brazil}
\author{Matthias Vojta}
\affiliation{Institut f\"ur Theoretische Physik and W\"urzburg-Dresden Cluster of Excellence ct.qmat, Technische Universit\"at Dresden, 01062 Dresden, Germany}

\date{\today}

\maketitle

\section{Spin stiffness}

We consider a spin-$S$ triangular-lattice Heisenberg model
\begin{equation}
    \begin{aligned}
        H &= J_1\sum_{\langle ij\rangle}\vec{S}_i\cdot\vec{S}_j
        + J_2\sum_{\llangle ij\rrangle}\vec{S}_i\cdot\vec{S}_j
    \end{aligned}
\end{equation}
in the regime $\alpha=J_2/J_1<1/8$ where the classical ground state is given by coplanar $120^\circ$ order. The spin stiffness is defined as the energy cost against a slow twist of the order parameter. The helical coplanar order introduces two stiffnesses, $\rho_s^\parallel$ for in-plane twists and $\rho_s^\perp$ for out-of-plane twists. For the nearest-neighbor model the classical stiffnesses have been quoted in Ref.~\onlinecite{dombre89}, and $1/S$ corrections have been calculated in Refs.~\onlinecite{chubukov94b,lecheminant95}.

As argued in the main text and shown below, bond defects cause in-plane distortions, such that $\rho_s^\parallel$ is the relevant stiffness. We restrict ourselves to the classical limit $S\to\infty$, where it is straightforward to include $J_2$ into the calculation, with the result
\begin{equation}
\label{eq:stiffclass}
\rho_s^\parallel =
    N_0^2 (J_1 - 6 J_2) \frac{\sqrt{3}}{2}\mathcal{A}
\end{equation}
where $\mathcal{A}=\sqrt{3}a^2/2$ is the unit cell area of the triangular lattice, with $a$ the lattice constant (which we set to unity unless explicitly written). Further, $N_0=S$ measures the order-parameter amplitude, i.e., the local spin expectation value in a rotated frame.
We note that continuum calculations often use a different convention for the stiffness prefactor, such that $\mathcal{A}$ is excluded from the above expression \cite{dombre89,chubukov94b}. We will use $\tilde\rho_s^\parallel = \rho_s^\parallel/\mathcal{A}$ to denote this in-plane ``continuum'' stiffness.

Evidently, positive $J_2$ reduces the stiffness compared to the nearest-neighbor model, consistent with increasing frustration. Notably, the stiffness \eqref{eq:stiffclass} remains finite at $J_2/J_1=1/8$ where the $120^\circ$ order gives way to a four-sublattice state, consistent with the (classical) transition being of first order.


\section{Single bond defect}

Here we provide some details concerning the physics of a single defect bond. To our knowledge, its dipolar nature has been first pointed out in Ref.~\onlinecite{syrom15}.

\subsection{Spin texture via linear response}
\label{sec:singresp}

As explained in the main text, the spin texture induced by a bond defect can be understood as the response to a local dipolar transverse field which acts to rotate the two spins on the defect bond away from their clean-limit configuration. This is most efficiently calculated in a local frame where the clean-limit state is uniform.

To this end we start from a coplanar state in the $x-y$ plane and rotate the spins such that all point along the $\hat{x}$ axis. The in-plane rotation due to the dipolar field involves only the $y$ components of the  spins, i.e., the static susceptibility $\chi^\parallel(\vec{q}) = \llangle S_{\vec{q}}^y; S^y_{-\vec{q}} \rrangle$ where $\llangle A; B\rrangle = \int_0^\beta d\tau \langle A(\tau) B \rangle$ with $\beta$ the inverse temperature.
For a system with underlying SU(2) spin symmetry, this susceptibility is governed by the in-plane Goldstone modes and follows $\chi^\parallel(\vec{q})= N_0^2/(\rho_s^\parallel q^2)$ for small momenta \cite{ssbook}.
For a weak defect of magnitude $\delta J $ on the bond $(i,j)=(0,1)$, the
transverse response at a separate site $i$ is given by
\begin{align}
    \langle S_i^\perp \rangle = \sum_{j=0}^1 h^\perp
    \beta_j \llangle S_i^y;
    S_j^y \rrangle
\label{resplatt}
\end{align}
where $\beta_j=(-1)^j$ is the dipolar form factor of the perturbation, and $h^\perp = \lambda (\delta J) N_0$ measures the locally transverse field  due to the bond defect. $\lambda = \sin \Theta_{\rm NN}$ is a geometrical factor, with $\Theta_{\rm NN}$ the angle between neighboring spins in the unperturbed state, $\lambda = \sqrt{3}/2$ for $120^\circ$ order.
Using the Fourier-transformed susceptibility in Eq.~\eqref{resplatt}, the long-distance piece of the response reads
\begin{align}
        \langle S_i^\perp\rangle
        = \lambda (\delta J) N_0\mathcal{A}
        \int \frac{d^d q}{(2\pi)^d}
        \left(i\hat{e}\cdot\vec{q}
        \right)\chi^\parallel(\vec{q})
        e^{i\vec q\cdot\vec r_i}
\label{master_resp}
\end{align}
where $\hat{e}$ is the directed defect bond vector and $d$ the spatial dimensionality.
Inserting the Goldstone-mode form for $\chi^\parallel(\vec{q})$ gives
\begin{align}
        \langle S_i^\perp\rangle
        = \delta J \frac{\lambda N_0^3}
        {\tilde\rho^\parallel_s}
        \int \frac{d^d q}{(2\pi)^d}
        \frac{i \hat{e}\cdot\vec{q}} {q^2}
        e^{i\vec q\cdot\vec r_i}\,.
\end{align}
Evaluating the integral and using $\delta\Theta(\vec r_i) = \langle S_i^\perp\rangle/N_0$ we finally find
\begin{equation}
    \begin{aligned}
        \delta\Theta(\vec r)
        &= \delta J \frac{\lambda N_0^2}
        {\tilde\rho_s^\parallel}\,
        \frac{\Gamma\left(d/2\right)}
        {2\pi^{d-1}\Gamma\left(2-\frac{d}{2}\right)}\,
        \frac{\hat{e} \cdot \vec{r}}{r^d}
    \end{aligned}
\label{eq:tr_resp}
\end{equation}
which describes a $d$-dimensional dipolar spin texture. Eq.~\eqref{eq:tr_resp} corresponds to Eq.~(3) of the main text, with the prefactor $\kappa=\lambda/(2\pi^{d-1}) \Gamma(d/2)/\Gamma(2-d/2)$, and $\Gamma(z) = \int_0^\infty dx x^{z-1} e^{-x}$ is the gamma function.

While the above derivation applies generally to a bond defect in a non-collinear magnet in $d$ space dimensions, the classical-limit response of the triangular-lattice Heisenberg model can also be calculated using spin-wave theory. The coefficients of the normal and anomalous pieces of the bilinear magnon Hamiltonian are \cite{deutscher93}
\begin{align}
  A_{\vec{q}}/S &= 3J_1 - 6J_2
        +\frac{3J_1}{2}\gamma_1(\vec q)
        +6 J_2  \gamma_2(\vec q), \\
  B_{\vec{q}}/S &= \frac{9J_1}{2}\gamma_1(\vec q)
\end{align}
with the geometric factors for the first-neighbor and second-neighbor interactions on the triangular lattice
\begin{align}
  \gamma_1(\vec q) &=\frac{1}{3}\left(
        \cos q_xa + 2\cos\frac{q_xa}{2}
        \cos\frac{\sqrt{3}q_ya}{2}
        \right),\\
  \gamma_2(\vec q) &= \frac{1}{3}\left(
        \cos \sqrt{3}q_ya + 2\cos\frac{3q_xa}{2}
        \cos\frac{\sqrt{3}q_ya}{2}
        \right).
\end{align}
The spin-wave energies are $\w_q^2=A^2_{\vec q} - B^2_{\vec q}$ while the in-plane susceptibility of the ordered state is given by \cite{deutscher93}
\begin{align}
\label{eq:chitriang}
    \chi^\parallel(\vec{q}) = \frac{S}{A_{\vec q} - B_{\vec q}}\,.
\end{align}
Expanding Eq.~\eqref{eq:chitriang} for small $q$ yields $1/\chi^\parallel(\vec{q}) = (3/4) (qa)^2 (J_1-6 J_2)$. Using $N_0=S$, valid in the classical limit, the susceptibility can be brought into the form $\chi^\parallel(\vec{q})= N_0^2/(\rho_s^\parallel q^2)$, with the stiffness $\rho_s^\parallel$ as given in Eq.~\eqref{eq:stiffclass}.

Evaluating \eqref{master_resp} using Eq.~\eqref{eq:chitriang}, the dipolar spin texture on the triangular lattice in the semiclassical limit reads explicitly
\begin{equation}
\delta\Theta(\vec r) = \frac{\delta J}{2\pi (J_1-6 J_2)}
        \frac{\hat{e} \cdot \vec{r}}{r^2}.
\end{equation}
This expression describes the data in Fig.~2(b) of the main paper, where $\delta J=-J_1/10$, with an accuracy better than 5\%.

\subsection{Uniform impurity moment}

\begin{figure}
\includegraphics[width=0.5\columnwidth]{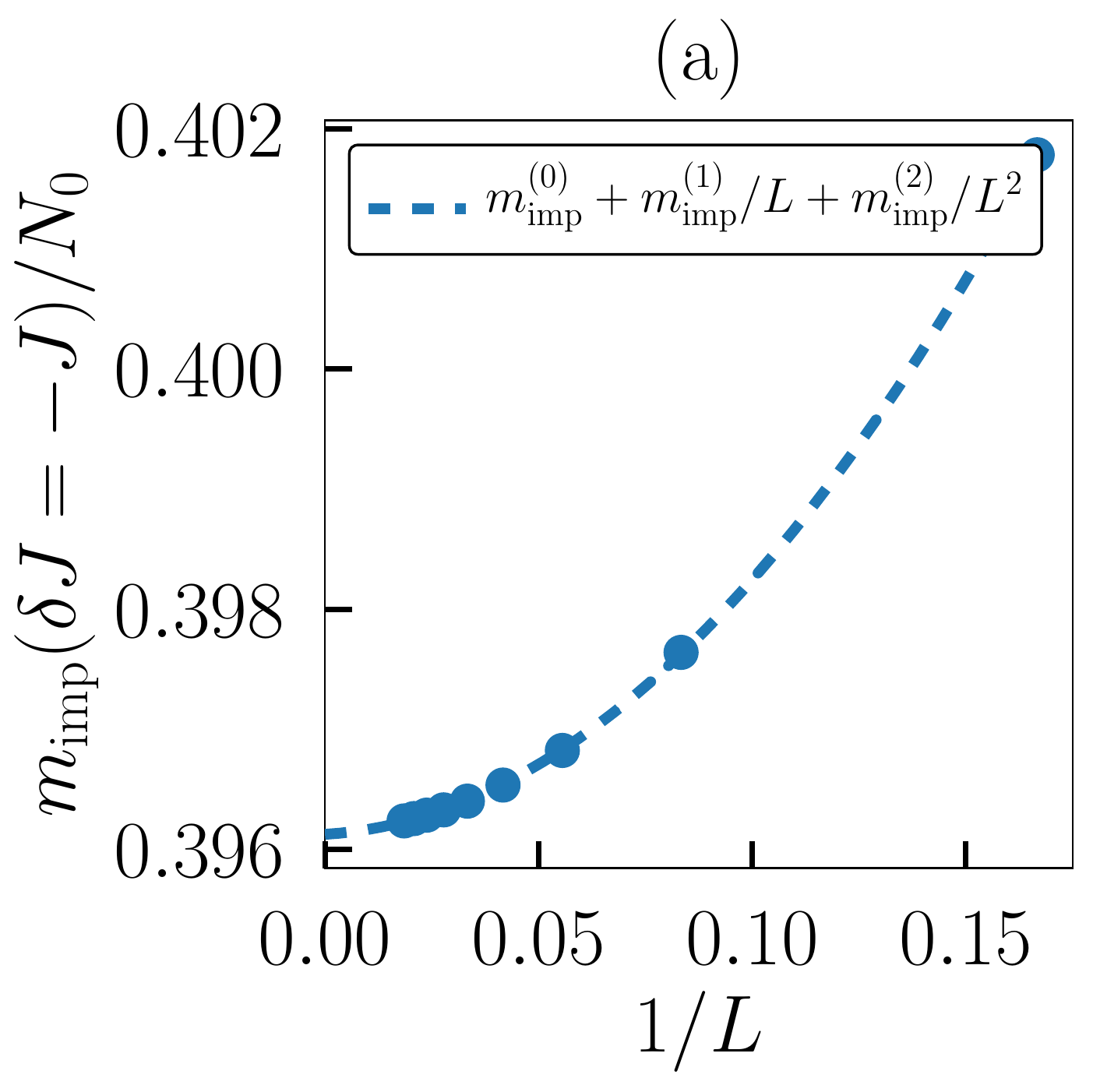}%
\includegraphics[width=0.5\columnwidth]{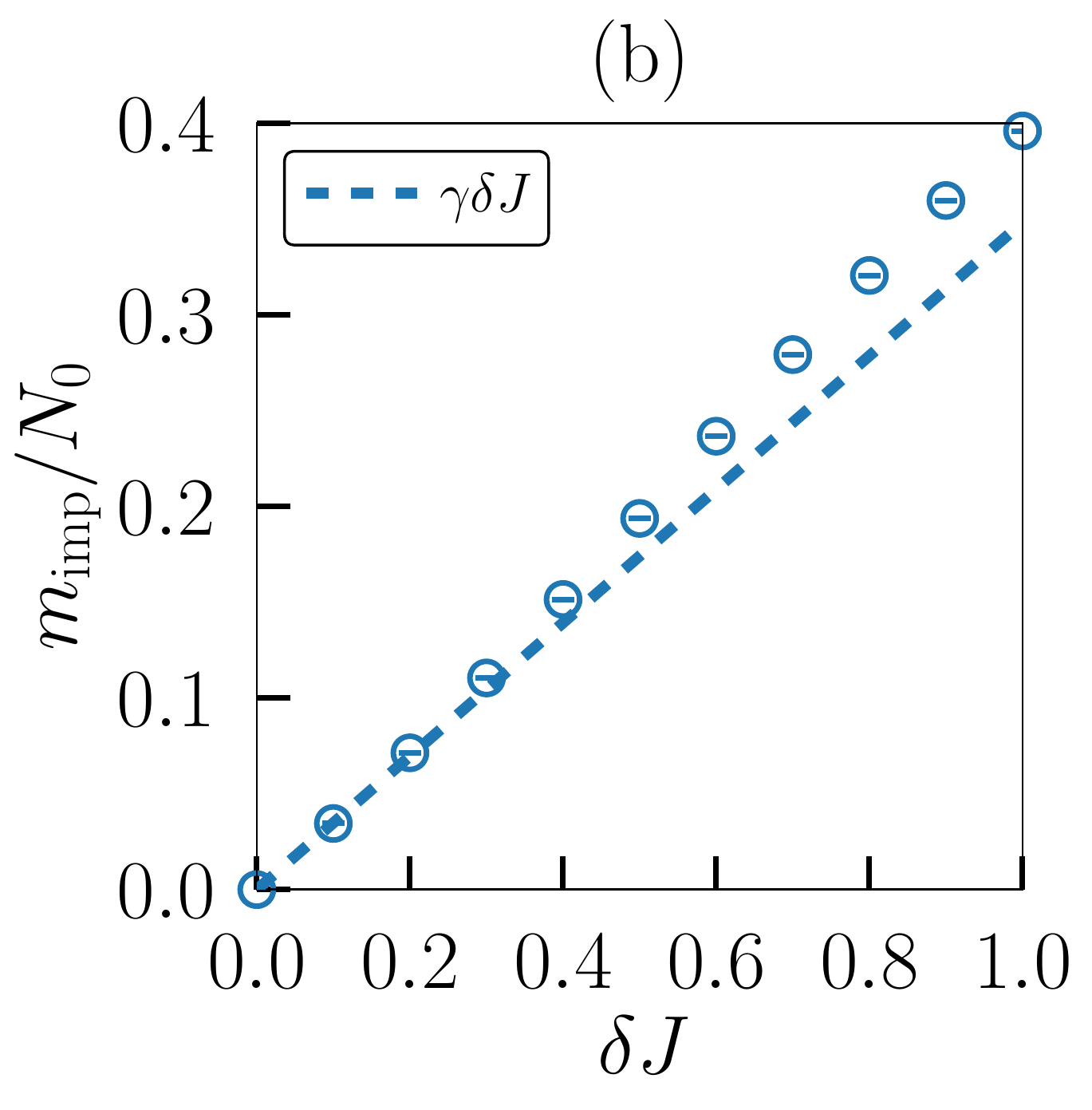}
\vspace*{-4pt}
\caption{Uniform moment generated due to bond defect $\delta J$.
(a) Finite-size scaling of the uniform moment for an absent bond at the center of the lattice.
(b) Growth of the uniform moment with increasing defect amplitude $\delta J$. For small values of $\delta J$ the relationship is linear.
}
\label{fig:uni_mo}
\end{figure}

In the presence of a single bond defect, the bulk spin moments do not exactly cancel each other, such that the resulting state has a non-vanishing uniform magnetization $\mimp$ of order unity (i.e. not proportional to system size). We have determined $\mimp$ numerically from the ground-state spin configurations on finite clusters; a corresponding finite-size scaling analysis of $\mimp$ is shown in Fig. \ref{fig:uni_mo}(a).
The impurity moment $\mimp$ can be expected to be proportional to $\delta J$ for small $\delta J$:
\begin{equation}
    \mimp= \eta N_0\delta J \,.
\end{equation}
For the classical triangular-lattice Heisenberg model with $J_2=0$ we numerically find $\eta = 0.35$, see Fig. \ref{fig:uni_mo}(b). For larger $\delta J$ corrections beyond linear response appear; for $\delta J=-J_1$ we have $\mimp/N_0 = 0.396$.

\section{Destruction of long-range order: Linear response}

Here we derive the heuristic criterion for the destruction of magnetic long-range order in dimensions $d\leq 2$ due to bond defects. The argument starts from helical LRO and shows that bond defects lead to divergent fluctuations of this order.

In the context of the triangular lattice, we assume weak bond disorder, i.e., spatially fluctuating nearest-neighbor couplings $J_{1,ij}$ drawn independently from a distribution with mean value $\overline{J_{1,ij}}=J_1$ and second moment $\overline{\delta J_{ij}^2}=\Delta^2$ where $\delta J_{ij} = J_{1,ij} - J_1$. Weak disorder in the $J_2$ couplings does not affect the $120^\circ$ state; thus we set $J_{2,ij} \equiv J_2$. More generally, our treatment is valid for any type of defects which produce a dipolar texture.

Weak disorder corresponds to $\Delta\ll J_1$; in this limit the influence of the defects can be treated perturbatively as in Sec.~\ref{sec:singresp}. As a result, we deal with a linear superposition of dipolar textures. Generalizing Eq.~\eqref{master_resp} yields
\begin{align}
\left\langle S_{l}^{\perp}\right\rangle &= \lambda N_0 \mathcal{A}
\sum_{\langle ij\rangle}
\int \frac{d^d q}{(2\pi)^d}
\left(i\vec{d}_{ij}\cdot\vec{q}\right)\chi^\parallel(\vec{q})
e^{i\vec q\cdot\vec r_{l,ij}}
\end{align}
where the sum runs over all (defect) bonds, $\vec{d}_{ij}=\hat{e}_{ij}\delta J_{ij}$ parameterizes the effective dipole strength on the bond $\hat{e}_{ij}$, and $\vec{r}_{l,ij}$ is the vector connecting site $l$ and the center of this bond.
After disorder averaging, this response is zero, but its fluctuations are finite:
\begin{equation}
        \overline{\langle {S_l^\perp}^2\rangle} =
        \lambda^2\Delta^2N_0^2\mathcal{A}^2
        \sum_{\hat{e}_\alpha}
        \int \frac{d^dq}{(2\pi)^d}
        \left(
        \hat{e}_\alpha\cdot\vec{q} \right)^2
        \chi^\parallel(\vec{q})^2
\label{auto_corr}
\end{equation}
where the $\hat{e}_\alpha$ represent the three bond orientations of the triangular lattice. We have $\sum_{\hat{e}_\alpha}(\hat{e}_\alpha\cdot\vec{q})^2= \zeta q^2$, with $\zeta = 3/2$ for the triangular lattice.
Inserting again the clean-limit Goldstone-mode form of $\chi^\parallel(\vec{q})$ results in
\begin{equation}
        \overline{\langle {S_l^\perp}^2\rangle} =
        \zeta\lambda^2\Delta^2\frac{N_0^6}{(\tilde\rho_s^\parallel)^2}
        \int \frac{d^dq}{(2\pi)^d}
    \frac{1}{q^2}\,.
\label{auto_corr2}
\end{equation}
This is equivalent to Eq.~(5) of the main text, with the prefactor $\tilde\kappa=2\pi^{d/2} \zeta\kappa^2/\Gamma(d/2)$. The integral in Eq.~\eqref{auto_corr2} is infrared divergent for $d\leq 2$, i.e., the random arrangement of dipoles destroys the assumed magnetic LRO.


\subsection{Two-dimensional SU(2)-symmetric systems}

According to Eq.~\eqref{auto_corr2}, $d=2$ is the marginal dimension for bond disorder in non-collinear Heisenberg antiferromagnets. Assuming that the integral is cutoff by a finite correlation length $\xi$, the stability criterion $\overline{\langle {S_{l}^{\perp}}^2\rangle } \lesssim N_0^2$ translates into an exponential dependence of the $T=0$ correlation length on the disorder level:
\begin{equation}
\ln \xi \propto \frac{(\tilde\rho_s^\parallel)^2}{\Delta^2 N_0^4}\,.
\label{corr_expr}
\end{equation}
We re-iterate that $\tilde\rho_s^\parallel$ and $N_0$ appearing here are characteristics of the clean system.

At finite $T$ and in $d=2$, the Mermin-Wagner theorem dictates that LRO is destroyed in the clean system due to thermal fluctuations. The resulting paramagnetic state has a magnetic correlation length which is exponentially large at low temperature, $\xi \sim e^{2\pi\tilde\rho_s/T}$ \cite{chakravarty89,azaria92}.
Comparing with the expression of the $T=0$ correlation length in the bond-disordered system \eqref{corr_expr} suggests the existence of a crossover temperature $T^\ast$, such quenched disorder (thermal fluctuations) dominate below (above) $T^\ast$, respectively. $T^\ast$ scales quadratically with the disorder level,
\begin{equation}
T^* \propto \Delta^2 N_0^4 / \tilde\rho_s \,.
\end{equation}
Here, the distinction between $\rho_s^\parallel$ and $\rho_s^\perp$ (which differ by a factor of two in the classical limit) has been ignored.

\subsection{Layered systems}

\begin{figure}
\begin{centering}
\includegraphics[width=0.9\columnwidth]{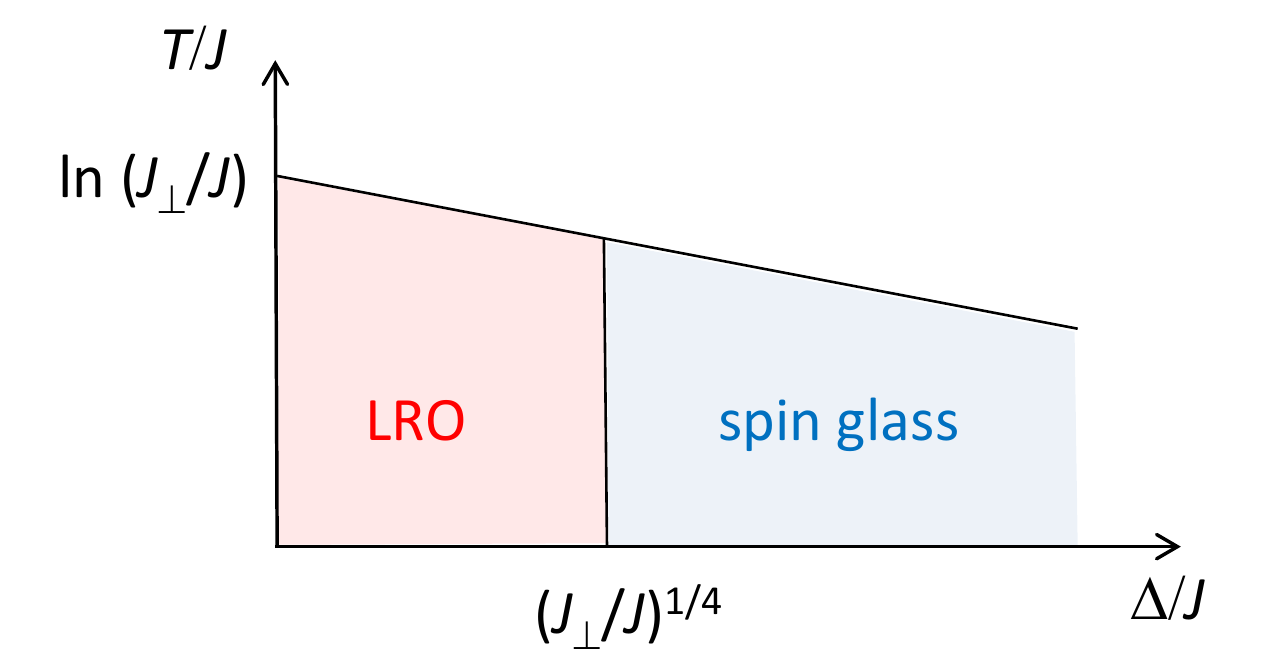}
\end{centering}
\caption{
Schematic phase diagram for quasi-2D non-collinear Heisenberg antiferromagnets with quenched bond disorder as function of disorder strength $\Delta$ and temperature $T$. LRO survives up to a critical strength of disorder which scales as $J_\perp^{1/4}$ where $J_\perp$ is the magnetic interlayer coupling.
}
\label{fig:pd_jp}
\end{figure}

Layered compounds realize quasi-two-dimensional systems with finite inter-layer coupling. For Heisenberg magnets, a small inter-layer exchange coupling $J_\perp$ stabilizes magnetic order even at finite temperatures: The corresponding ordering temperature scales logarithmically with $J_\perp$.
In the ordered state, the susceptibility can now be written
\begin{equation}
\chi^\parallel(\vec{q},q_\perp) =
\frac{N_0^2}{\rho_s}\frac{1}{q^2+\varepsilon q_\perp^2}
\end{equation}
where $q_\perp$ is the momentum perpendicular to the layers, and $\varepsilon\ll 1$ parameterizes the inter-layer coupling, e.g., $\varepsilon \sim J_\perp/J_1$ for stacked triangular lattices.

Inserting this into the fluctuation expression \eqref{auto_corr}, we see that the integral divergence is now cut off, i.e., the momentum integral scales as $\sqrt{\varepsilon}$. Then, the stability condition implies that LRO is stable provided that
\begin{equation}
    \varepsilon \gtrsim
    \frac{\Delta^4 N_0^8}
        {(\tilde\rho_s^\parallel)^4},
\end{equation}
i.e. the critical level of disorder required to destroy LRO scales as $(J_\perp/J_1)^{1/4}$. The resulting phase diagram is shown in Fig.~\ref{fig:pd_jp}.


\subsection{Broken SU(2)}

If SU(2) spin symmetry is broken at the Hamiltonian level, some or all of the Goldstone modes of the ordered phase acquire a gap. For the case of non-collinear order with gapped in-plane Goldstone mode, the spin texture discussed in Sec.~\ref{sec:singresp} will decay exponentially instead of algebraically. The susceptibility can be approximated as
\begin{equation}
\chi^\parallel(\vec{q}) = \frac{N_0^2}{\rho_s}\frac{1}{q^2+(m/c)^2}
\end{equation}
with $c\propto J$ a velocity. The mode gap $m$ removes the infrared divergence of the integral \eqref{auto_corr} in two space dimensions. As a result, LRO is stable for small disorder, i.e., if
\begin{equation}
\ln (m/J) \gtrsim \frac{(\tilde\rho_s^\parallel)^2}{\Delta^2 N_0^4}\,.
\end{equation}

For an easy-plane system with U(1) symmetry at the Hamiltonian level, the perturbative considerations for weak disorder apply as in the SU(2)-symmetric case, and LRO is destroyed for arbitrarily small disorder. However, the phenomenology of the resulting phase is more complicated and will be explored in future work \cite{dey19b}.


\section{Destruction of long-range order: Renormalization-group analysis}

In this section, we sketch an alternative route to derive the destruction of non-collinear LRO by weak bond disorder. This is based on a field-theoretic formulation of semiclassical order and a renormalization-group treatment of the effect of bond disorder.

\subsection{Field theory}

The long-distance properties of a coplanar magnet can be described by an order-parameter field $R(\tau, \vec x)$ where $R\in$\,SO(3) is a rotation matrix, defined on the lattice relative to a fixed set of vectors $\vec{N}_i$ in spin space which describe the spiral ordering \cite{dombre89,azaria93},
\begin{equation}
    \begin{aligned}
        \vec{S}_i=R_i\cdot\vec{N}_i,
    \end{aligned}
\end{equation}
for the $120^\circ$ order the $N_i$ are three different vectors corresponding to the three sublattices. The coplanar order is symmetric under the transformation $R\rightarrow U R V^{-1} $, where $U\in$\,SO(3) is the global spin rotation symmetry and $V\in$\,O(2) is the continuum enlargement of the discrete space group of the planar lattice (which is $C_{3v}$ for triangular lattice, see Ref.~\onlinecite{azaria93} for details).
The effective low-energy action is a non-linear sigma model (NLSM) of the matrix field $R$, first appearing in Ref.~\onlinecite{dombre89}:
\begin{equation}
    \begin{aligned}
        \mathcal{S}=&
        -\frac{\tilde{\rho}}{4}
        \int d\tau d^dx
        \left[
            \frac{1}{c^2}
            \Tr\left(R^{-1}\partial_\tau R\right)^2
        +\Tr P\left(R^{-1}\partial_i R\right)^2
        \right]
    \end{aligned}
\end{equation}
where $\tilde{\rho}$ is the continuum stiffness \cite{convention} and $P$ is a projection operator on to the plane of the ordered state, satisfying $\left[P,V\right]=0$.
Calculations are simplified in the vielbein basis parametrization  $\omega^a_\mu t_a = R^{-1}\partial_\mu R$ \cite{dombre89,azaria93} where $\mu=0,\ldots,d$ is a space-time index and $t_a\in$\,Lie[SO(3)] are the three generators of SO(3) with normalization $\Tr\left[t_at_b\right] = -2\delta_{ab}$,
\begin{equation}
    \begin{aligned}
        \mathcal{S}=&
        \frac{\tilde{\rho}}{2c^2}
        \int d\tau d^dx
        \left[
            \omega_\tau^{a}(\tau,\vec{x})^2
        +c^2_a \omega_i^{a}(\tau,\vec{x})^2
        \right] \,.
    \end{aligned}\label{clean_act}
\end{equation}
For the triangular lattice antiferromagnet, the bare anisotropy
in the spin-wave velocities is given by
$
c^2_{x,y}
=c^2_z/2
=c^2/2
$
and the corresponding bare stiffnesses are
$\tilde{\rho}_{x,y}
=\tilde{\rho} c^2_{x,y}/c^2,
\tilde{\rho}_z
=\tilde{\rho} c^2_z/c^2
$
\cite{dombre89,chubukov94b}.

As detailed in Sec.~\ref{sec:singresp}, a bond defect acts as a directed dipolar perturbation. It hence couples with a single gradient to the local order-parameter field:
\begin{equation}
    \begin{aligned}
        \delta J_{ij}
        \vec{S}_i\cdot\vec{S}_j
        &=\delta J_{ij}
        \vec{N}_i\cdot R^{-1}
        \left(
        \hat{e}_{ij}\cdot\grad
        \right)
        R\cdot\vec{N}_j\\
        &=\delta J_{ij}N_0^2
        \lambda
        \hat{e}_{ij}\cdot
        \vec{\omega}^z
    \end{aligned}
\end{equation}
where the $R$ and $\omega$ fields are taken at the defect position. For a finite defect concentration, disorder averaging over the random defect terms can be performed using the standard replica trick \cite{fisher85}, and we arrive at the action
\begin{equation}
    \begin{aligned}
        \mathcal{S}=&
        \frac{\eta_a}{2}
        \sum_r
        \int_{\tau,x}
        \left[
            \omega_\tau^{ra}(\tau,\vec{x})^2
        +\omega_i^{ra}(\tau,\vec{x})^2
        \right]\\
        &-\frac{\sigma\eta_z^2}{2}
        \sum_{rs}
        \int_{\tau,\tau',x} \
        \omega_i^{rz}(\tau,\vec{x})
        \omega_i^{sz}(\tau',\vec{x}).
    \end{aligned}\label{inter_act}
\end{equation}
Here are $r$ and $s$ are replica indices, the temporal axis was appropriately scaled \cite{chakravarty89}, and the bulk couplings $\eta_a=\tilde{\rho}_a/c_a$ were introduced. The strength of the disorder is encoded in $\sigma \propto \Delta^2$, more precisely $\sigma\tilde{\rho}_z^2 =\Delta^2 \zeta\lambda^2N_0^4a^2/\mathcal{A}$.

\subsection{Renormalization group and destruction of LRO}

While quenched disorder only couples to the in-plane mode in Eq.~\eqref{inter_act}, a coupling to the other low-energy modes is generated under renormalization. We therefore start with a more general action
\begin{align}
\mathcal{S} &=
        \frac{1}{2}
        \int_{\tau,\tau',x}
        K^{rs}_{a\mu}(\tau,\tau')
        \omega^{ra}_\mu(\tau,\vec{x})
        \omega^{sa}_\mu(\tau',\vec{x})\,, \notag \\
        K^{rs}_{a\mu}(\tau,\tau')
        &=\eta_a\delta(\tau-\tau')
        \delta^{rs}
        -\sigma_a\eta_a^2\delta_{\mu i}
\label{rpl_act}
\end{align}
where the bare disorder couplings are $\sigma^z \propto \Delta^2$ and $\sigma_{x,y}=0$. The tree-level scaling dimensions
of the couplings are respectively,
$\text{dim}\left[\eta_a\right] = d-1$ and
$\text{dim}\left[\sigma_a\right] = 2-d$,
confirming that quenched bond disorder is marginal in $d=2$ and relevant in $d<2$.

The fate of the ordered state can be studied using renormalization-group techniques. Following Polyakov \cite{polyakov-book}, we track the low-energy flow of the couplings by integrating out fast modes over momentum shells $(b\Lambda,\Lambda)$. We split the
replica flavored matrices $R^r$ into fast modes, $U^r=e^{t_a\varphi^{ra}}$ with $\varphi$ being spin-wave fields, and slow modes, $\tilde{R}^r$, such that $R^r=U^r\tilde{R}^r$.
Because we are ultimately interested in the replica symmetry unbroken classical modes $\tilde{R}^r=\tilde{R}$ \cite{fisher85}, the expansion in the vielbein basis is simplified,
\begin{equation}
    \begin{aligned}
        R^{r^{-1}}\partial_\mu R^r
        &=\tilde{\omega}^{ra}_\mu t_a
        +t_a
        \left[
            \partial_\mu\varphi^{ra}
            +\frac{1}{2}
            f_{abc}\partial_\mu
            \varphi^{rb}
            \varphi^{rc}
            +\dots
            \right]
    \end{aligned}
\end{equation}
where $f_{abc}$ are the structure constants of $SO(3)$, $[t_a,t_b]=f_{abc}t_c$. Plugging this back to the replica action \eqref{rpl_act} and integrating out the $\varphi$ fields, we determine the one-loop corrections to the bilinear $\omega$ terms. Demanding form invariance of this part of the action then yields the RG flow equations as
\begin{widetext}
\begin{equation}
    \begin{aligned}
        \beta(\eta_a) &=
        \frac{d\eta_a}
        {d\log b} =
        (1-d)\eta_a
        +
        \frac{R_{abc}}{8\pi}
        \left[
        \left(
        \frac{\eta^2_a}
        {\eta_b\eta_c}
        \right)
        +
        \frac{2\eta_a^2\sigma_c}{\eta_b}
        \right]
        \\
        \beta(\sigma_a)&=
        \frac{d\sigma_a}{d\log b}
        =(d-2)\sigma_a
        -\frac{R_{abc}}{8\pi}
        \left[
        \left(
        \frac{\sigma_b}{\eta_c}
        +\frac{2\eta_a\sigma_a}
        {\eta_b\eta_c}
        \right)
        +2(\sigma_b\sigma_c
        +\frac{2\eta_a\sigma_a\sigma_c}{\eta_b})
        \right]
    \end{aligned}\label{rg_flow}
\end{equation}
\end{widetext}
where $R_{abc}=\left(f_{abc}\right)^2$ is a symmetric traceless tensor, and we have restricted ourselves to the replica-diagonal limit. As announced, couplings of disorder to all spin-wave modes are generated under renormalization.

We now discuss an approximate solution of these RG equations. For $d$ close to 2, the $\eta_a$ are strongly relevant. As a result, the contributions to $\beta(\sigma_a)$ which are suppressed by a factor of $1/\eta$ are small and can be neglected. In $d=2$, this leaves us with $(-\sigma^2)$ terms in the $\beta$ functions of $\sigma_a$, underlining that disorder is marginally relevant.
We make no attempt to fully solve the coupled system of RG equations. Instead, we consider a simplified flow trajectory where all $\sigma_a$ grow in parallel, with relative factors taken from the structure of Eq.~\eqref{rg_flow}. Solving the resulting single flow equation, we obtain the correlation length as the scale where $\sigma$ reaches unity, with the result
\begin{equation}
\ln \frac{\xi^{\rm RG}}{\xi_\infty} = {\frac{2\pi\tilde{\rho}_z^2}
        {\mathcal{C}\zeta\lambda^2\Delta^2N_0^4a^2/\mathcal{A}}}
\label{xiRG}
\end{equation}
where $\mathcal{C}=2.3875$ for the triangular lattice. In Fig. 3(d) of the main text we make a comparison of this estimate against the numerically measured correlation lengths.

We note that a related conclusion, i.e., the absence of LRO in $d=2$ due to random dipolar perturbation, was reached in Refs.~\onlinecite{hasselmann04,cherepanov99} in the context of glassy physics in cuprate superconductors, earlier discussed in Ref.~\onlinecite{glazman90}. Although both our setting and technical treatment are different, with a non-collinear clean-limit state and a full consideration of its multiple Goldstone modes, our qualitative conclusion parallels that of Refs.~\onlinecite{hasselmann04,cherepanov99}.


\section{Numerical simulations}

To study the physics beyond the weak-disorder limit, we have performed large-scale numerical simulations for the classical $J_1$-$J_2$ triangular-lattice Heisenberg model. For finite lattices with $N=L^2$ sites and periodic boundary conditions, we have generated disordered sets of couplings by drawing each $J$ independently either from a Gaussian distribution,
\begin{equation}
    P(J_{\alpha,ij}) = \frac{1}{\sqrt{2\pi\Delta_\alpha^2}}
    e^{-\frac{\left(J_{\alpha,ij}-J_\alpha\right)^2}
    {2\Delta_\alpha^2}}, \ \alpha=1,2
\end{equation}
or from a bimodal distribution,
\begin{equation}
    P(J_{\alpha,ij}) =
    \begin{cases}
        \frac{1}{2} &\text{ for } J_{\alpha,ij}
        = J_\alpha - \Delta_\alpha\\
        \frac{1}{2} &\text{ for } J_{\alpha,ij}
        = J_\alpha + \Delta_\alpha
    \end{cases}, \ \alpha = 1,2
\end{equation}
where we simulate for $\Delta_2/\Delta_1=J_2/J_1$. In the data presented throughout this paper we abbreviate the disorder strength as $\Delta\equiv\Delta_1$.

At $T=0$ we have employed an iterative classical energy minimization scheme \cite{walker77} to find a locally stable configuration; per disorder realization we have used up to $N_{\rm init} \sim 400$ different initializations with varying degree of disorder and used the converged state with globally lowest energy to calculate observables. Averaging has been performed over $N_{\rm avg} \sim 1000$ realizations of disorder.
The maximum system size was limited by our ability to reliably find a low-energy state; for large $L$ we found this prohibitively expensive because a huge $N_{\rm init}$ is needed.

\begin{figure}
\begin{centering}
\includegraphics[width=0.5\columnwidth]{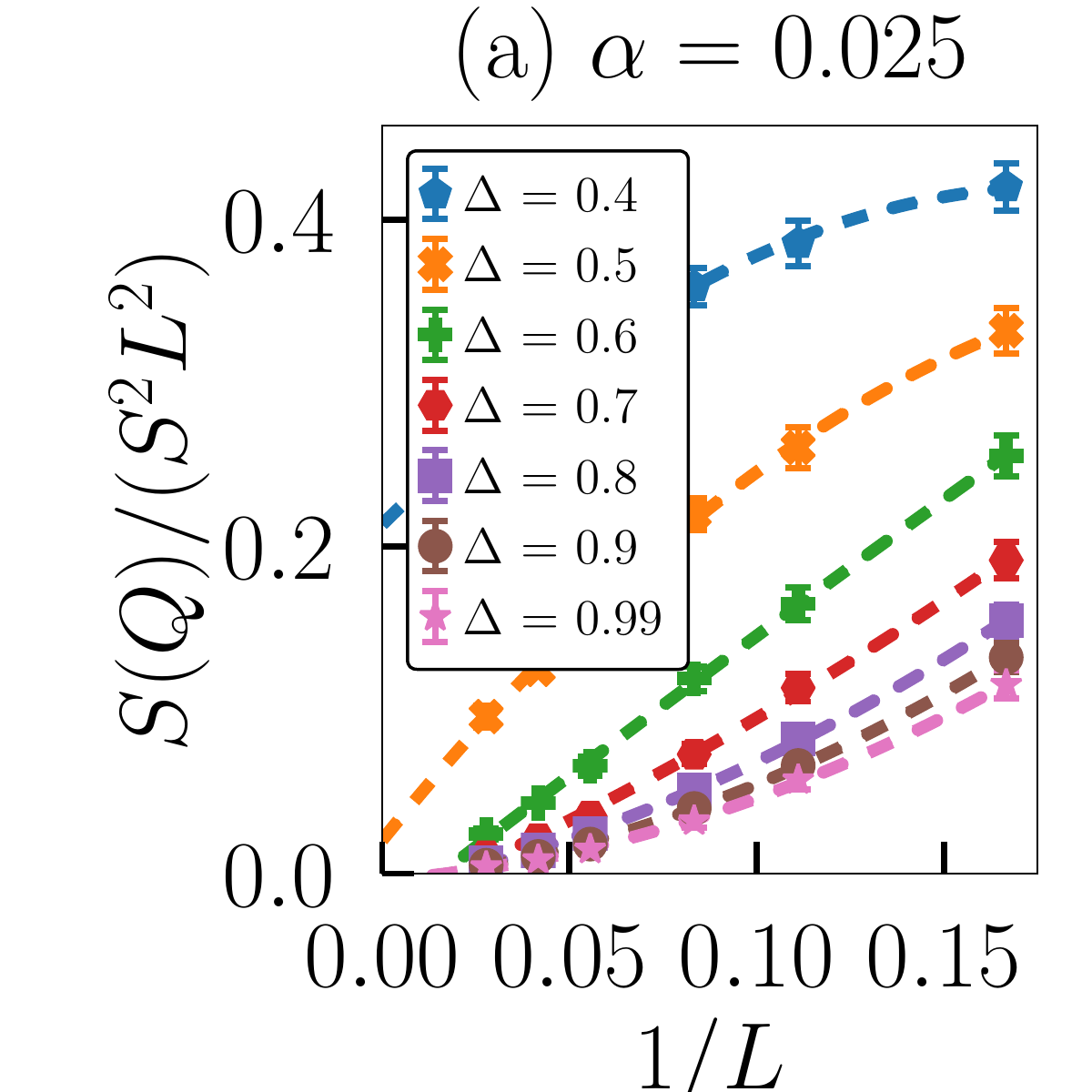}%
\includegraphics[width=0.5\columnwidth]{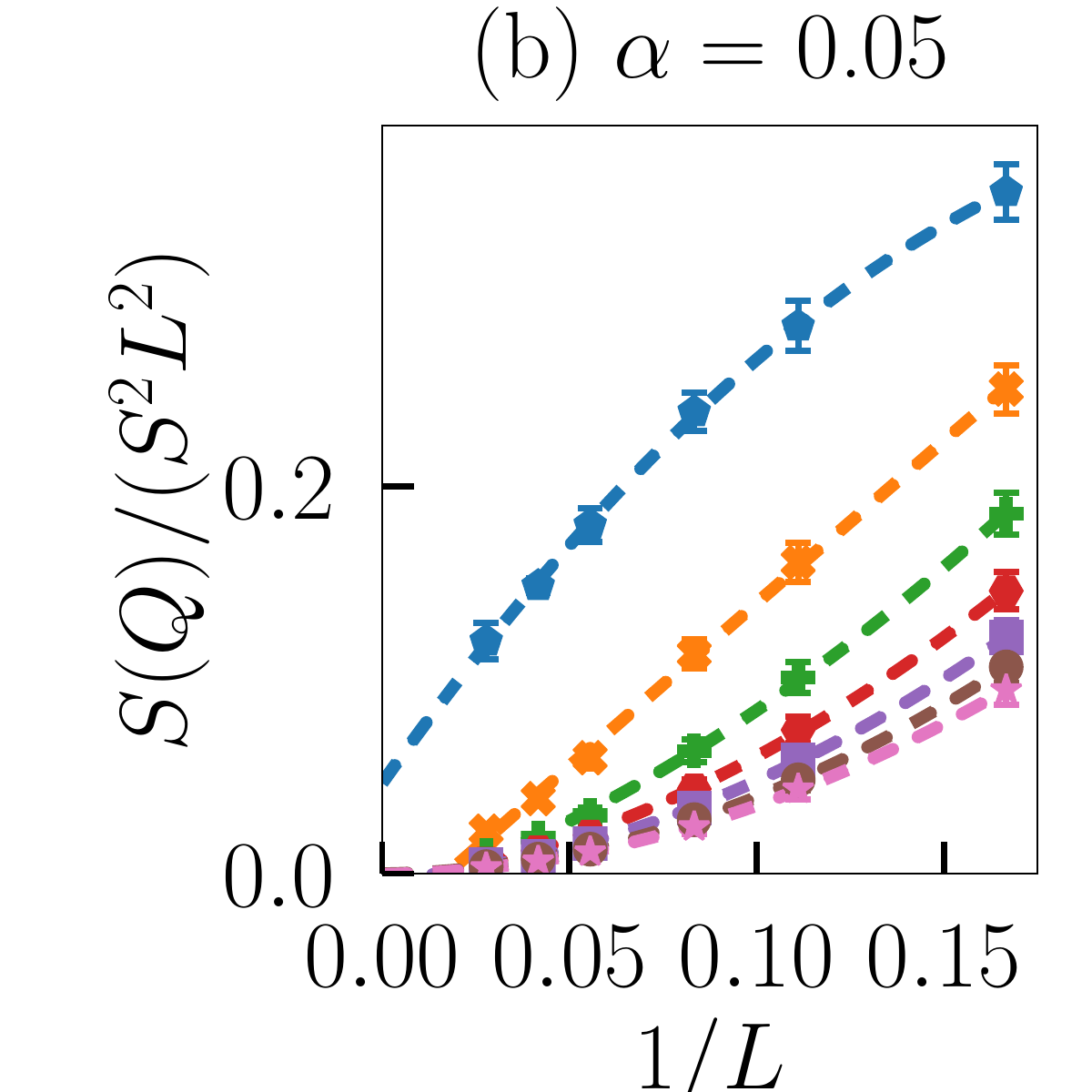}
\end{centering}
\caption{
Squared order parameter for the classical triangular-lattice Heisenberg model with Gaussian bond disorder as function of linear system size $L$ for different values of $\alpha=J_2/J_1$.
(a) $\alpha=0.025$,
(b) $\alpha=0.05$.
The dashed lines show a parabolic fit to the $1/L$ dependence. The change in curvature roughly coincides with the correlation length being comparable to the system size.
}
\label{fig:sq}
\end{figure}

\begin{figure*}
    \begin{centering}
\includegraphics[width=0.5\columnwidth]{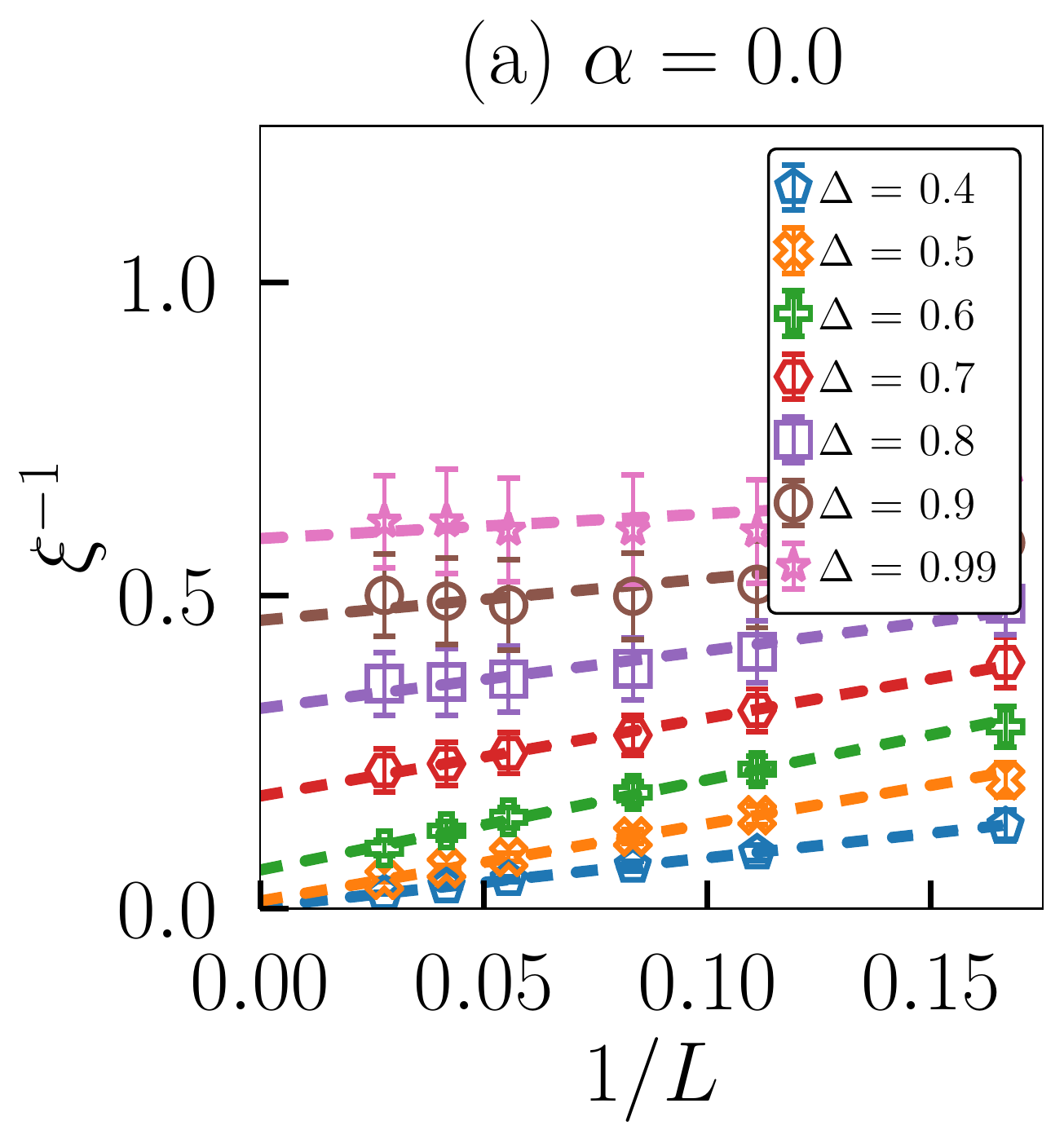}%
\includegraphics[width=0.5\columnwidth]{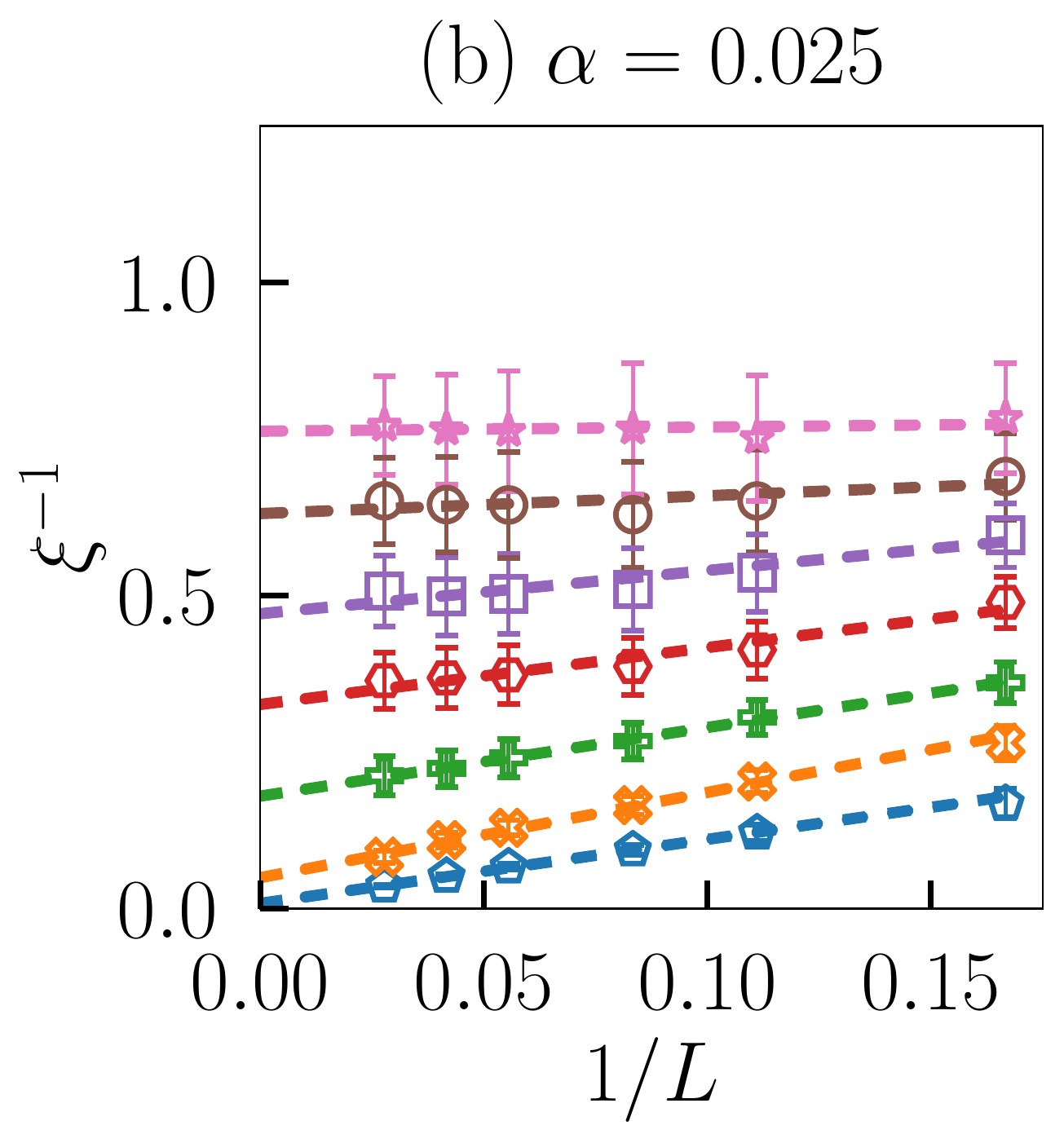}%
\includegraphics[width=0.5\columnwidth]{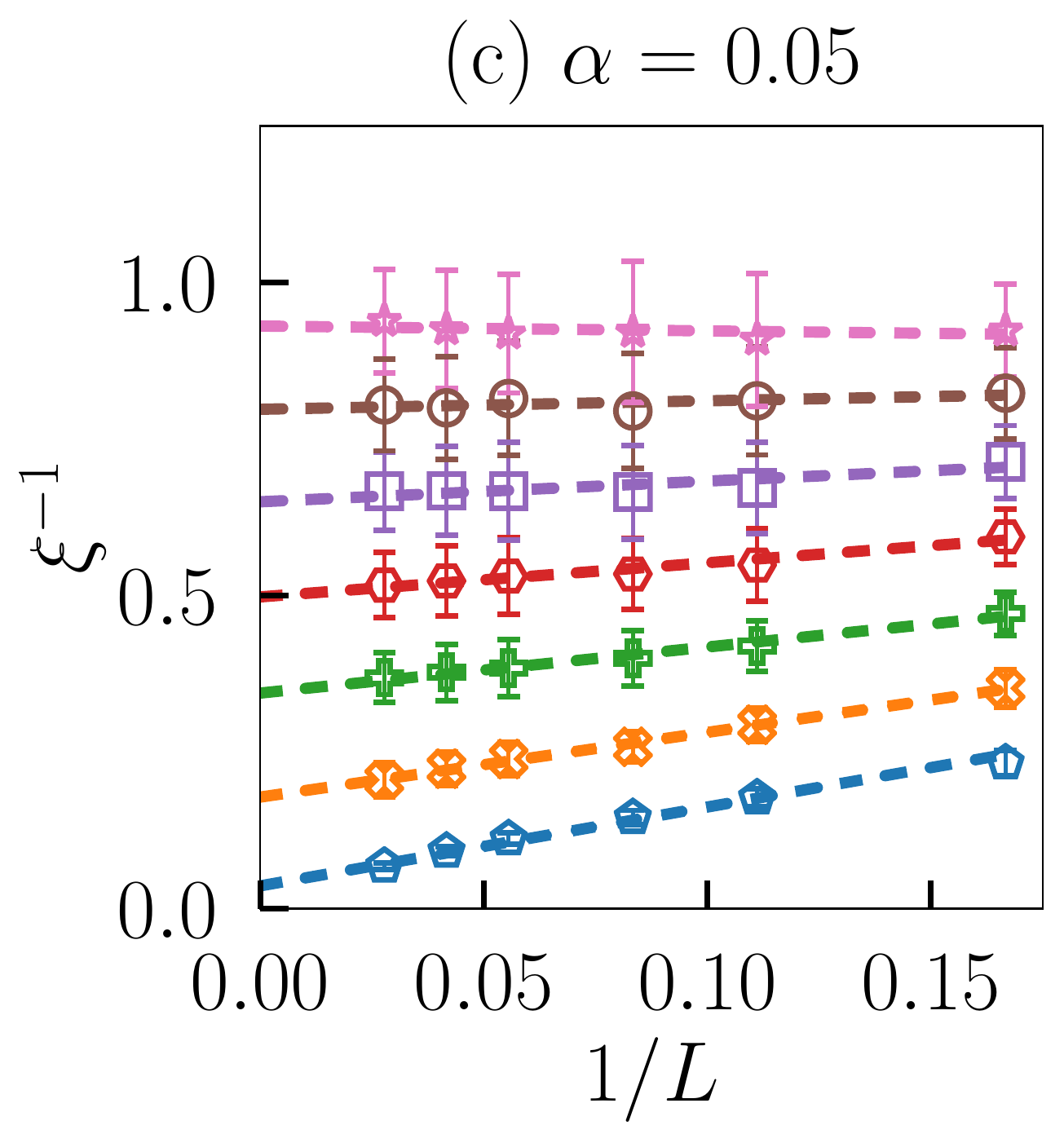}%
\includegraphics[width=0.5\columnwidth]{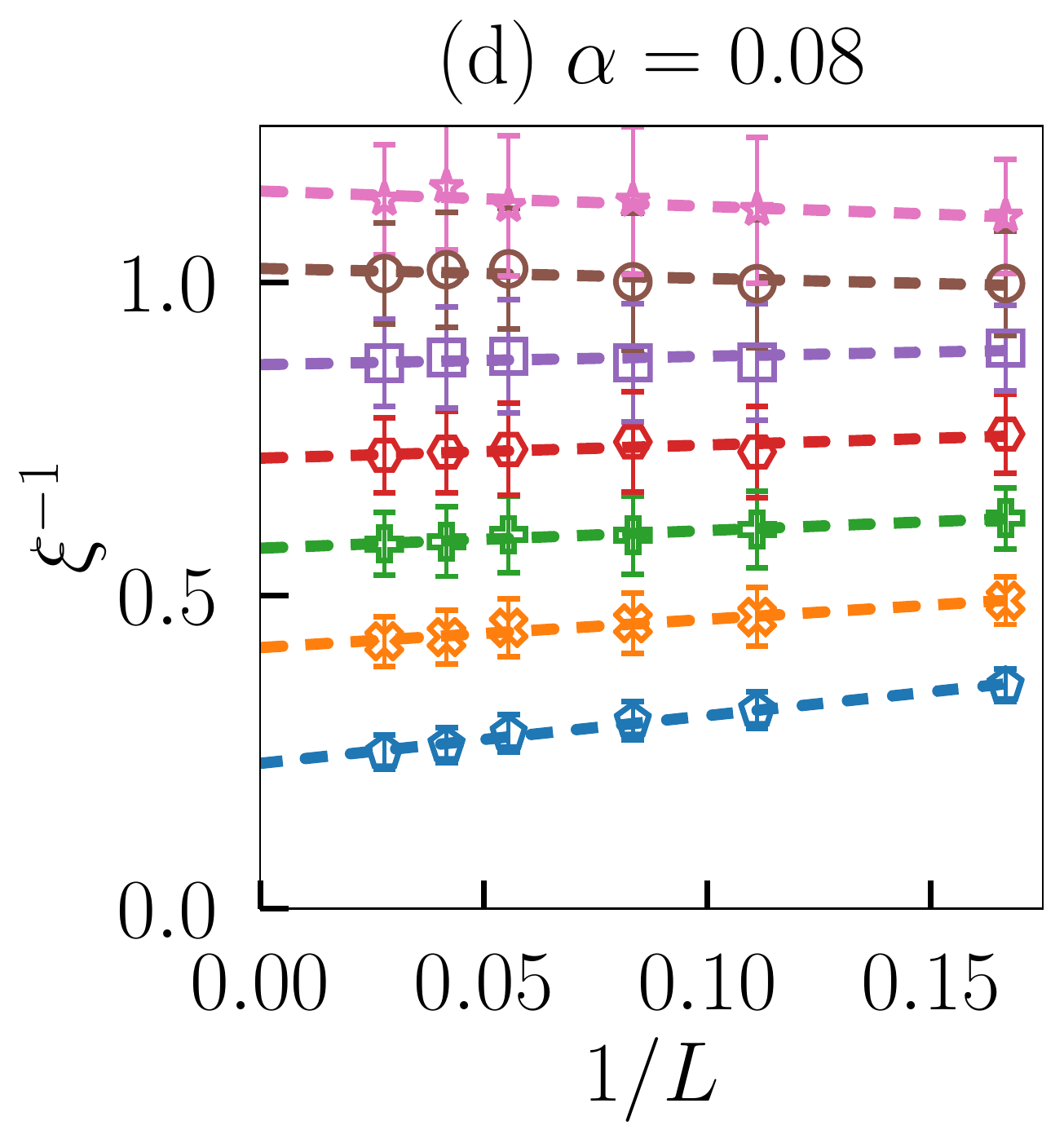}
    \end{centering}
\caption{
Finite-size scaling of the magnetic correlation length $\xi$ for the classical triangular-lattice Heisenberg model with Gaussian bond disorder for different values of $\alpha=J_2/J1$.
(a) $\alpha=0$,
(b) $\alpha=0.025$,
(c) $\alpha=0.05$,
(d) $\alpha=0.08$.
The dashed lines show linear fits of $1/\xi$ as function of $1/L$.
}
\label{fig:xifs1}
\end{figure*}
\begin{figure*}
    \begin{centering}
\includegraphics[width=0.5\columnwidth]{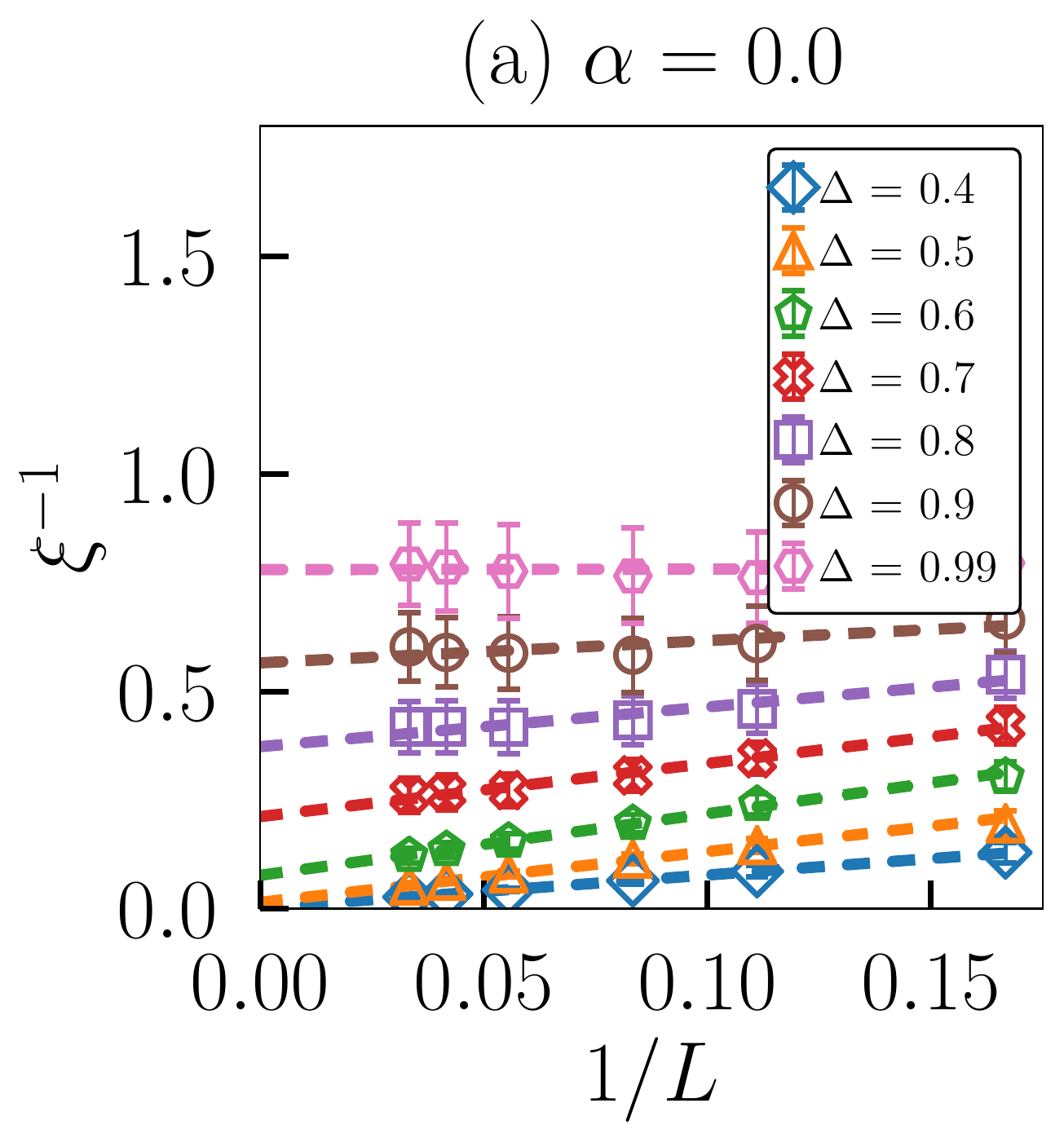}%
\includegraphics[width=0.5\columnwidth]{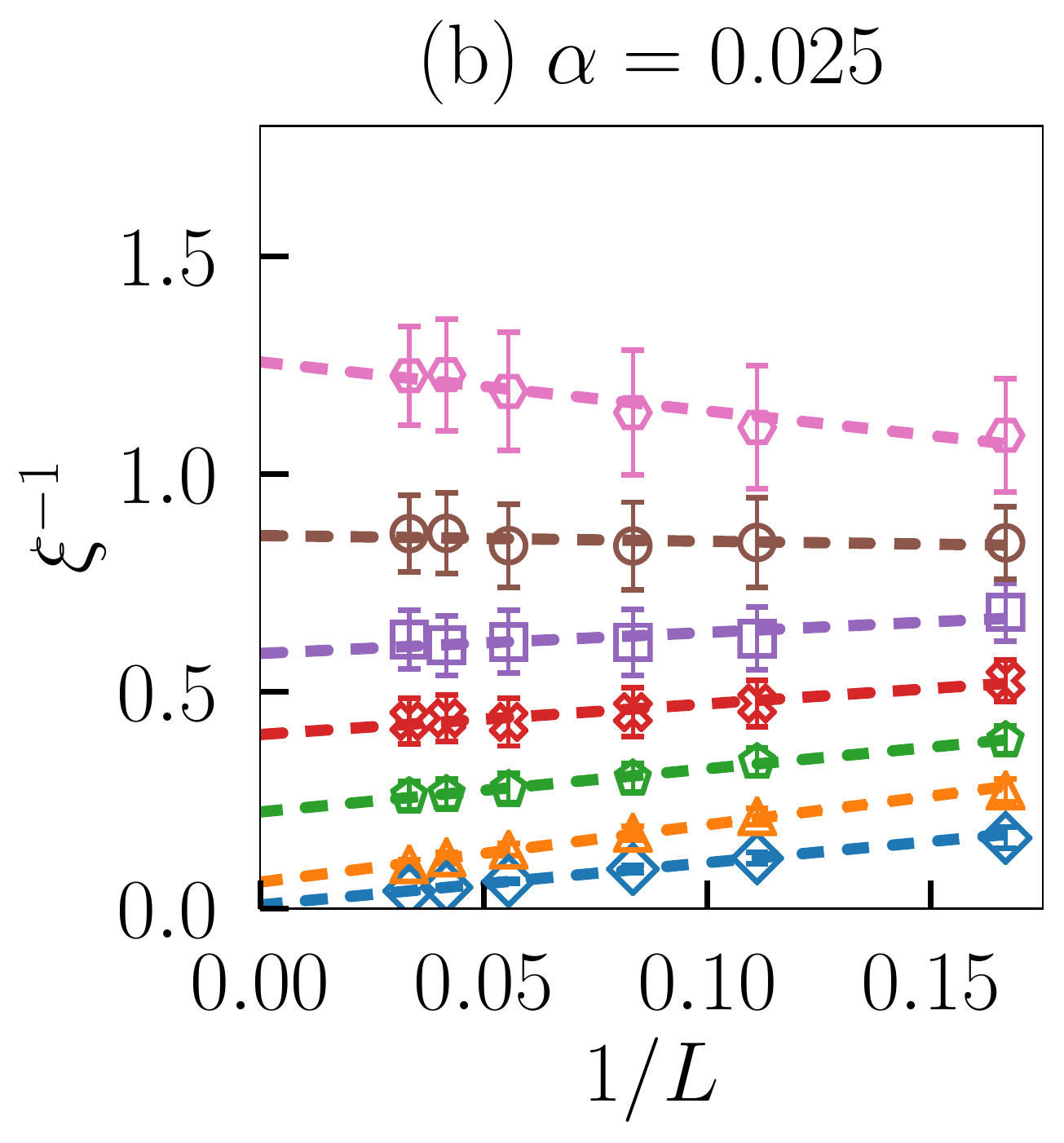}%
\includegraphics[width=0.5\columnwidth]{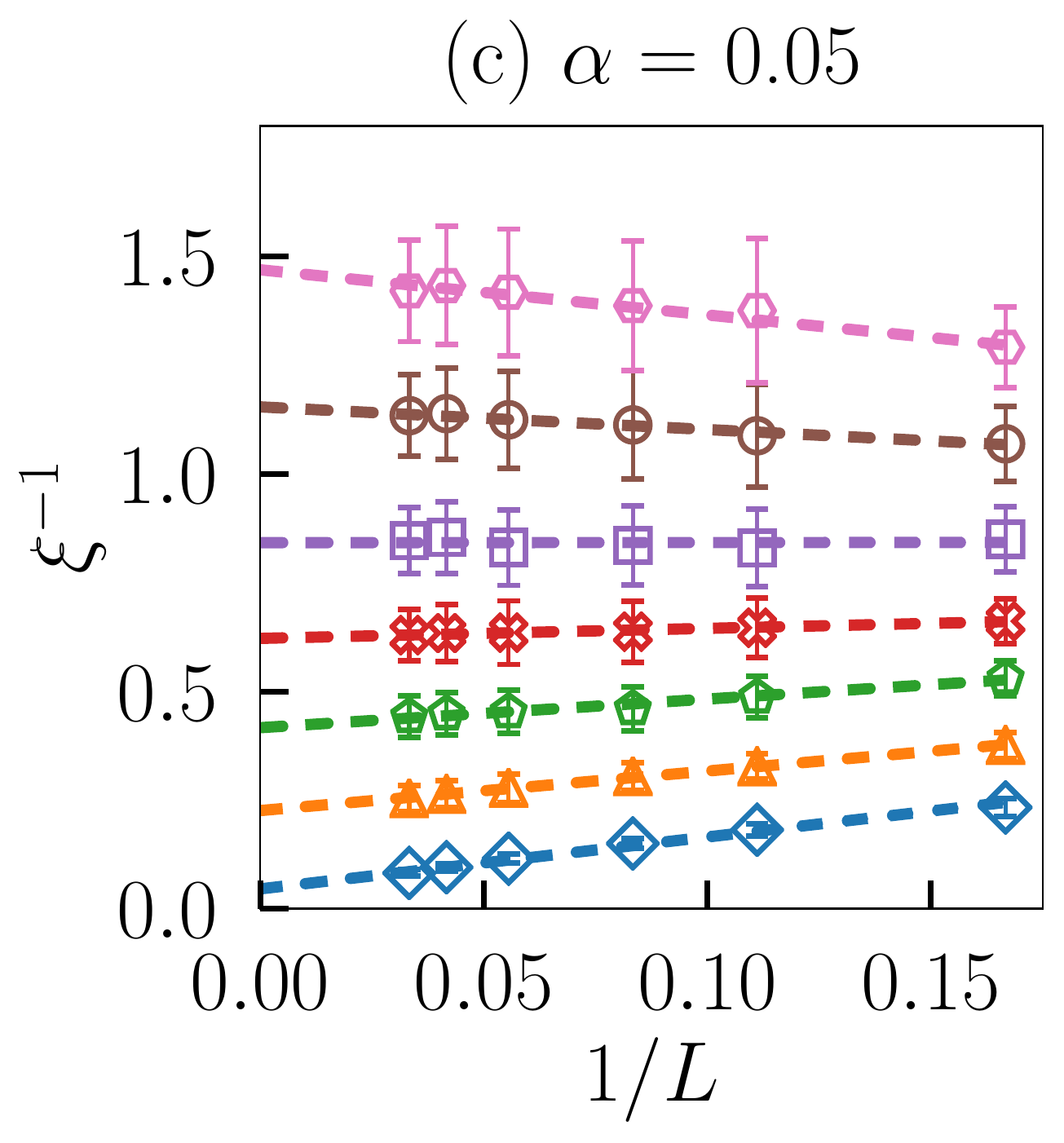}%
\includegraphics[width=0.5\columnwidth]{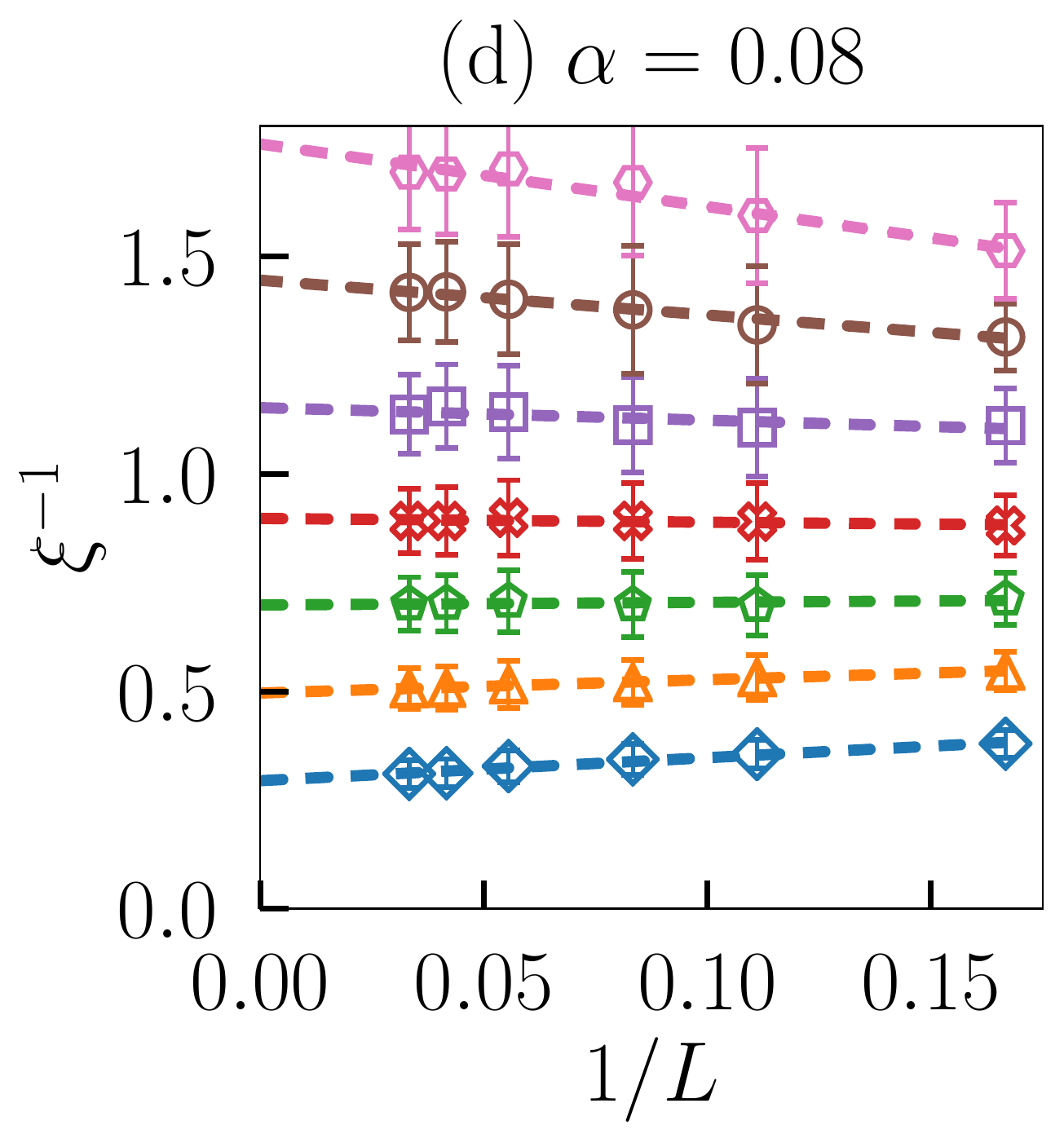}
    \end{centering}
\caption{
Same as Fig.~\ref{fig:xifs1}, but now for bimodal bond disorder.
}
\label{fig:xifs2}
\end{figure*}

In addition, we have performed finite-temperature Monte-Carlo simulations on lattices with $L^2$ sites. We employ three distinct types of MC moves: (a) single-site (restricted) Metropolis updates, (b) microcanonical steps~\cite{alonso96} and (c) parallel tempering~\cite{partemp}. Typically, we perform $5\times10^{5}$ MC sweeps for thermalization, followed by $5\times10^{5}$ sweeps to calculate thermal averages. In our implementation, after $10$ microcanonical sweeps we perform a Metropolis sweep followed by a parallel tempering update. For the restricted Metropolis step, we use a temperature-dependent selection window to ensure an average acceptance rate larger than $50\%$ at any given temperature. Moreover, we select our temperature grid such that a parallel tempering move has a success rate larger than $40\%$.
%

\subsection{Spin correlations: Additional results}

We have determined the disorder-averaged static structure factor
\begin{equation}
\overline{S(\vec q)} = \frac{1}{L^2} \sum_{ij} \overline{\langle\vec S_i \cdot \vec S_j\rangle} e^{i\vec q \cdot (\vec R_i-\vec R_j)}
\end{equation}
where $\langle\ldots\rangle$ and $\overline{\ldots}$ denote MC and disorder average, respectively. For a state with magnetic LRO at wavevector $\vec Q$, the value of $m^2 = \overline{S(\vec Q)}/N$ corresponds to the square of the order parameter $m$ in the thermodynamic limit.
In Fig.~\ref{fig:sq} we show the finite-size scaling of $m^2$ for different $\alpha$ and $\Delta$ for Gaussian disorder. The data clearly point to the absence of LRO for the parameters with $\alpha=0.025$, $\Delta\geq 0.6$ and $\alpha=0.05$, $\Delta\geq0.5$. For smaller $\Delta$  larger systems would be needed to draw a sharp conclusion.

The spin correlation length $\xi$ has been determined from $S(\vec q)$ in two ways, either by using the values of $\overline{S(\vec Q)}$ and the closest wavevector nearby, or by identifying $1/\xi$ as the full-width-half-maximum (FWHM) of $\overline{S(\vec q)}$ by fitting it to a Lorentzian as function of $|\vec q-\vec Q|$ along a cut in momentum space. Both protocols gave consistent results, with the FWHM method being more robust at large disorder. Hence we show $\xi$ obtained from FWHM here and in the main paper.

Figs.~\ref{fig:xifs1} and \ref{fig:xifs2} summarize the finite-size scaling of the $T=0$ data for $\xi$ for different $\alpha$ and $\Delta$, both for Gaussian and bimodal disorder. The error bars reflect the statistical uncertainty from the disorder average. In general, a linear dependence of $1/\xi$ on $1/L$ appears to fit the data. However, given that the largest system size is $L=36$, values of $\xi$ beyond 20 have to be considered unreliable. Therefore, data for smaller $\Delta$ are not shown.

Taken together, the data in Figs.~\ref{fig:sq}, \ref{fig:xifs1}, and \ref{fig:xifs2} are consistent with the LRO being destroyed by bond disorder in favor of a short-range ordered state; for small disorder the resulting spin correlation lengths are too large to be accessible in our finite-size simulations.

\begin{figure}
\includegraphics[width=0.5\columnwidth]{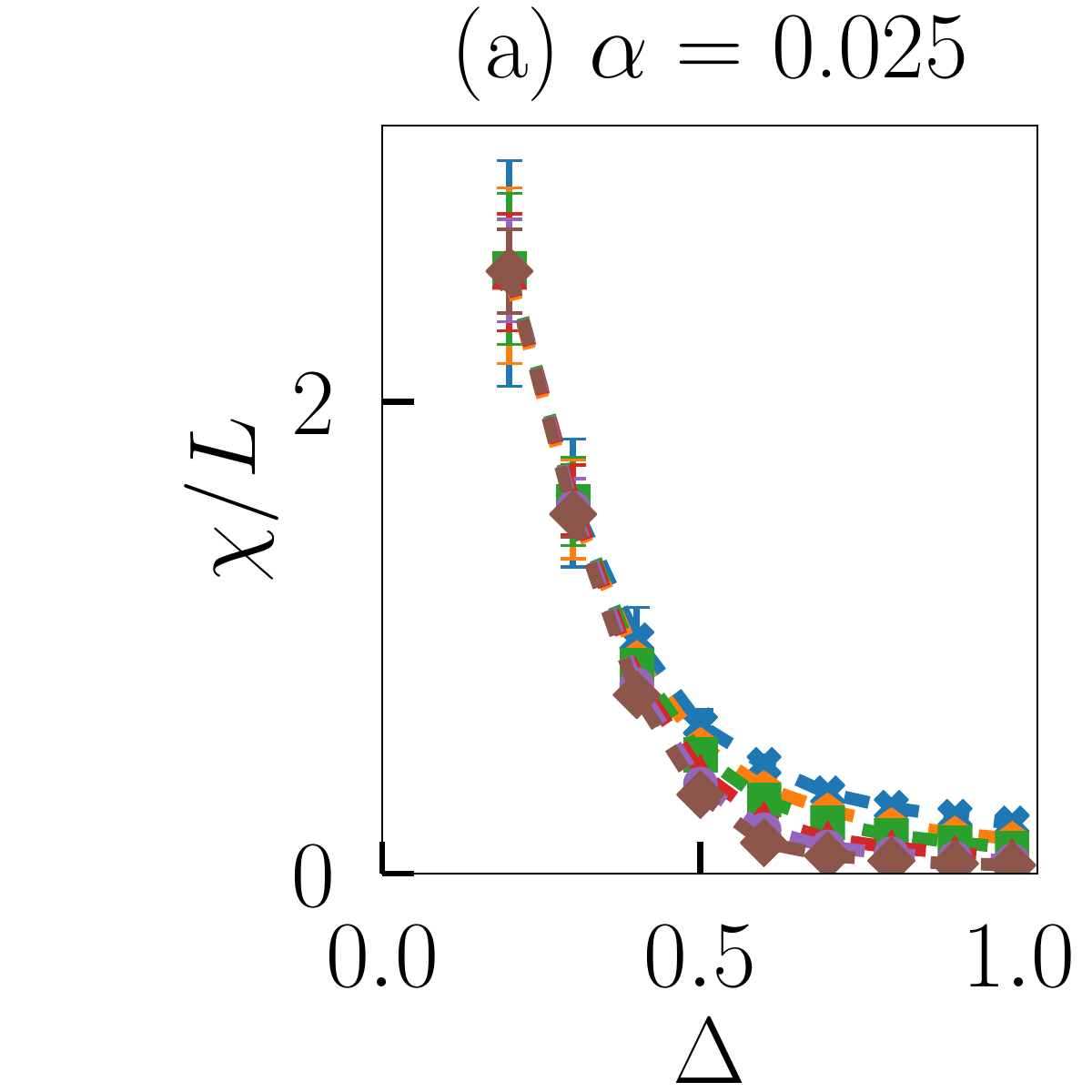}%
\includegraphics[width=0.5\columnwidth]{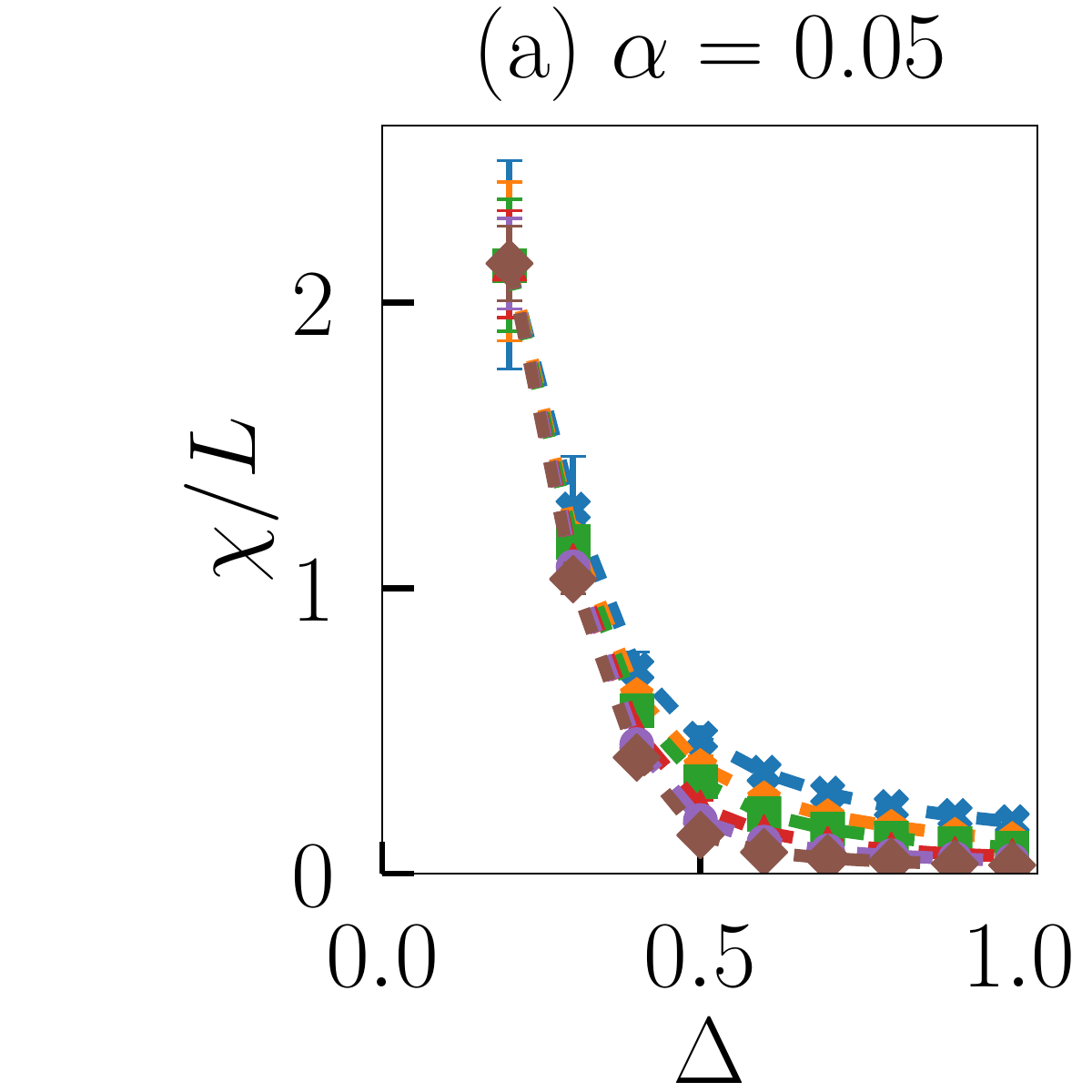}
\caption{
Magnetic correlation length $\xi/L$, now plotted as function of disorder strength $\Delta$ for Gaussian disorder and different $L$ for (a) $\alpha=0.025$ and (b) $\alpha=0.05$. No crossing points indicative of a critical disorder strength $\Delta$ are seen.
}
\label{fig:crossxi}
\end{figure}

\begin{figure}
\includegraphics[width=0.5\columnwidth]{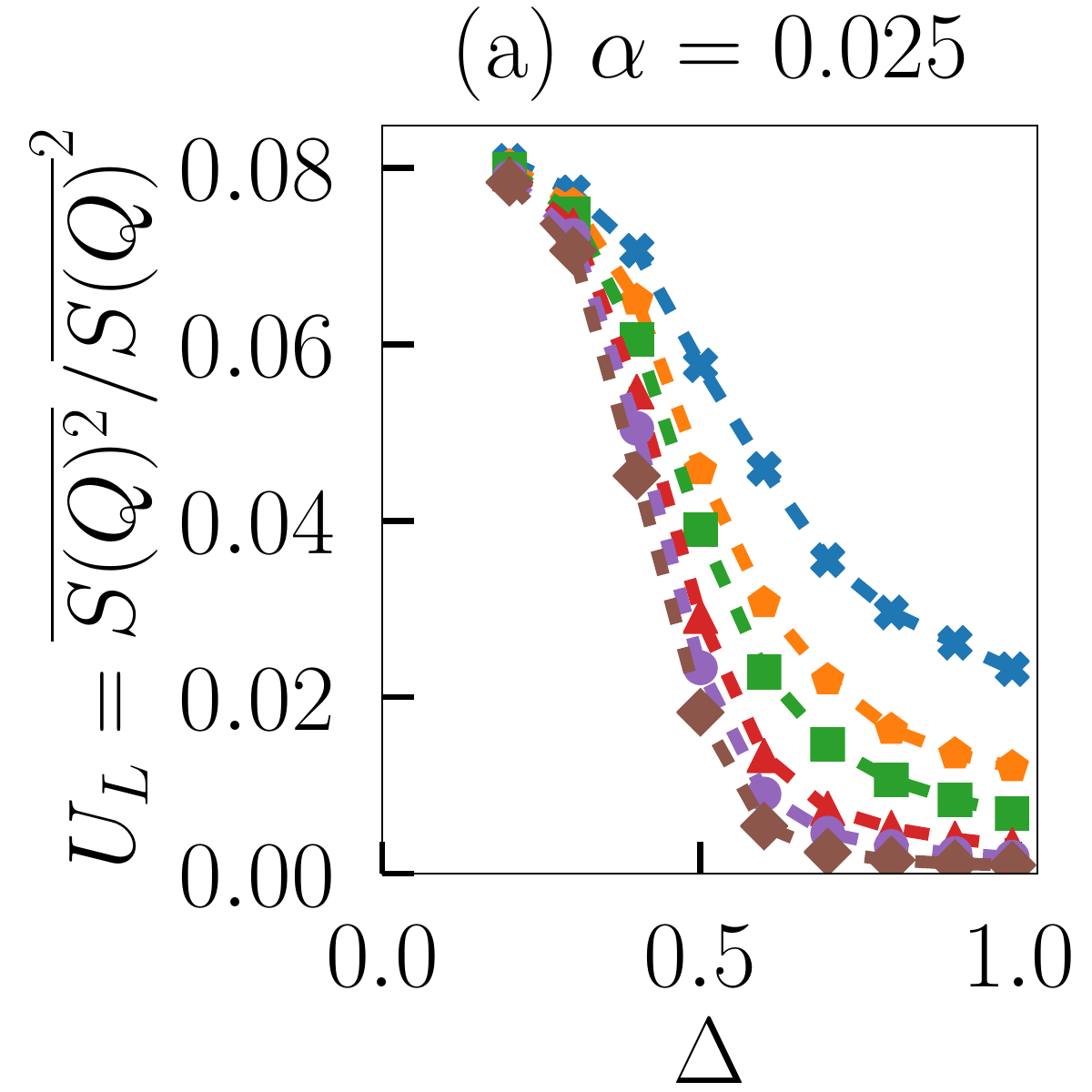}%
\includegraphics[width=0.5\columnwidth]{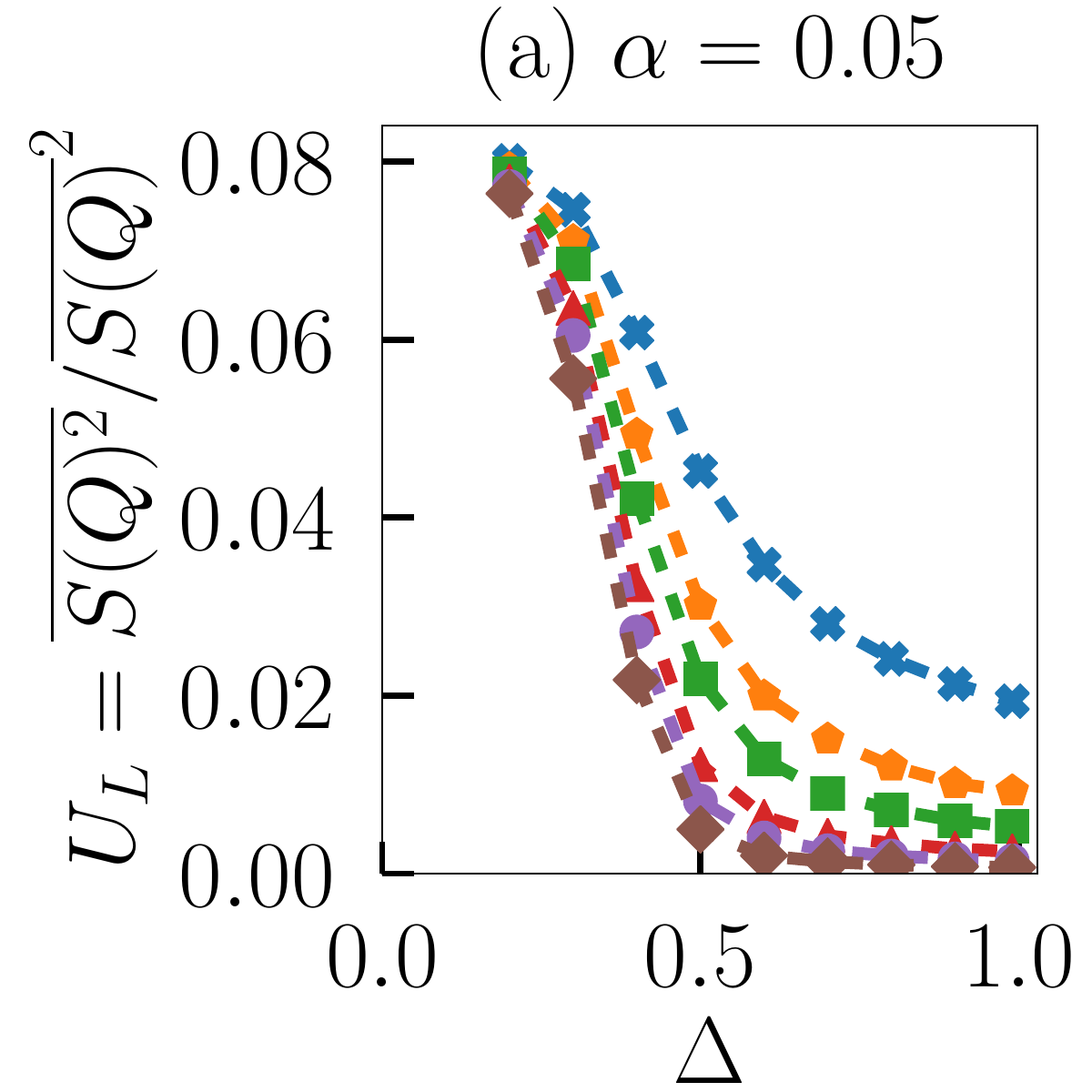}
\caption{
Binder parameter $U_L$ as function of disorder strength $\Delta$ for Gaussian disorder and different $L$ for (a) $\alpha=0.025$ and (b) $\alpha=0.05$. Again, no crossing points indicative of a critical disorder strength $\Delta$ are visible.
}
\label{fig:crossbinder}
\end{figure}

We have tested the data for signatures of critical behavior at $T=0$, by looking for a crossing point of the $\Delta$ dependence of the correlation length $\xi/L$ as function of $L$: Such a crossing point would signal a scale-invariant system. Remarkably, we find no crossing points at all, Fig.~\ref{fig:crossxi}. We have repeated this analysis with the Binder parameter $U_L \equiv \overline{S(\vec Q)^2}/\overline{S(\vec Q)}^2$, with the same result, Fig.~\ref{fig:crossbinder}. This strongly suggests the absence of a critical $\Delta$ and consequently the absence of LRO for any level of bond disorder.

\begin{figure}
\includegraphics[width=0.5\columnwidth]{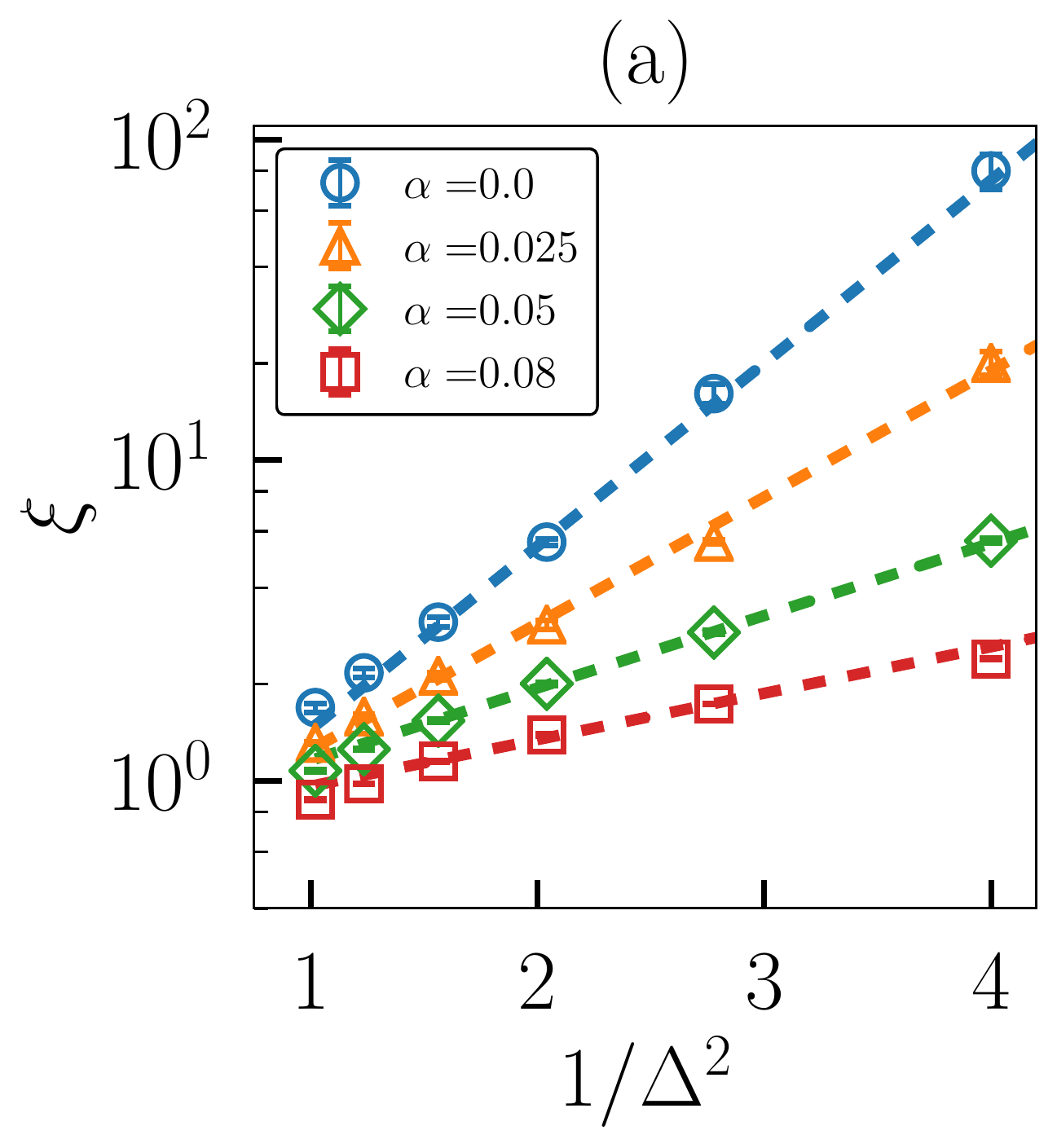}%
\includegraphics[width=0.5\columnwidth]{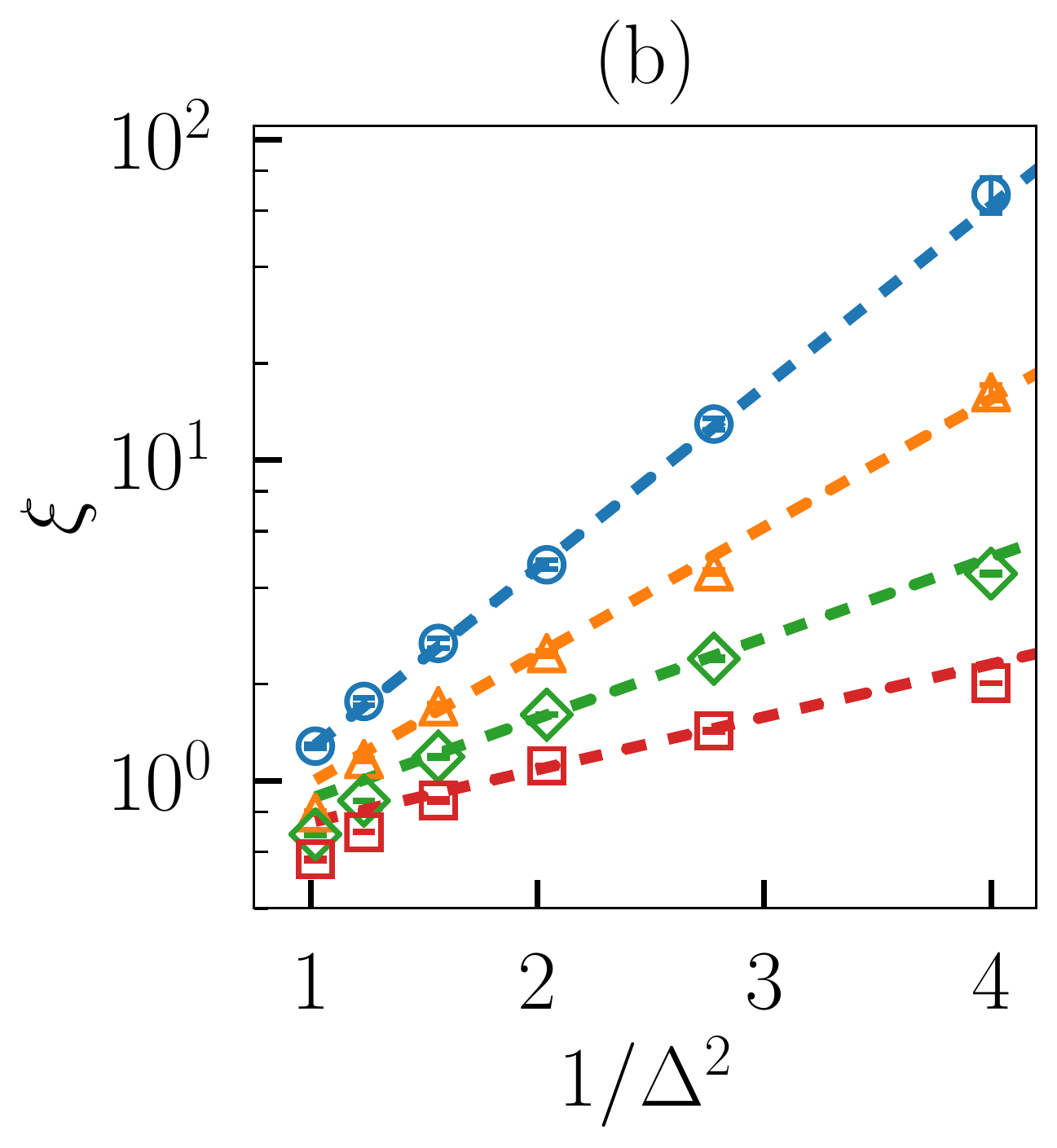}
\vspace*{-4pt}
\caption{
Magnetic correlation length $\xi$ as function of disorder strength $\Delta$, plotted as $\ln\xi$ vs. $1/\Delta^2$, for (a) Gaussian and (b) bimodal disorder and different values of $\alpha=J_2/J_1$. The dashed lines are linear fits; the linear dependence of $\ln\xi$ on $1/\Delta^2$ is consistent with Eq.~\eqref{corr_expr}.
}
\label{fig:xidelta}
\end{figure}

The exponential dependence of $\xi$, extrapolated to infinite system size as shown in Figs.~\ref{fig:xifs1} and \ref{fig:xifs2}, on the disorder strength, Eq.~\eqref{corr_expr}, is confirmed in Fig.~\ref{fig:xidelta}. Extrapolating this dependence to small $\Delta$ we obtain, e.g., for $\Delta=0.2$ and $\alpha=0$ the extremely large value $\xi \approx 10^{13}$ while for $\alpha=0.08$ we have $\xi\approx 10^5$. Both experiments and numerical simulations will therefore detect the destruction of LRO only if disorder exceeds a resolution-dependent threshold.

\subsection{Glassiness}

To confirm the spin-glass character of the disordered system, we have employed the finite-$T$ MC simulations to study the the spin-glass susceptibility \cite{fischer}
\begin{equation}
\chi_{\rm SG}(\vec{q})=N\sum_{\alpha,\beta} \overline{\langle
\left|q^{\alpha,\beta}\left(\vec{q}\right)\right|^{2}\rangle}
\end{equation}
where
\begin{equation}
q^{\alpha,\beta}\left(\vec{q}\right)=
\frac{1}{N}\sum_{i}S_{i}^{\alpha\left(1\right)}S_{i}^{\beta\left(2\right)}e^{i\vec{q}\cdot\vec{r}_{i}}
\end{equation}
is the momentum-dependent spin-glass order parameter. Here $\alpha$ and $\beta$ are spin components, $^{(1,2)}$ denote identical copies of the system (``replicas'') containing
the disorder configuration. The spin-glass correlation length $\xi_{\rm SG}$ is obtained from the behavior of $\chi_{\rm SG}(\vec{q})$ near ${\vec q}=0$, and the spin-glass order parameter is $m_{\rm SG}^2=\chi_{\rm SG}(0)$.
We note that $m_{\rm SG}$ is trivially non-zero at $T=0$ for the classical system at hand, as it measures autocorrelations of the spins over time, and in this limit the system freezes into a static spin configuration.

\begin{figure}
\includegraphics[width=\columnwidth]{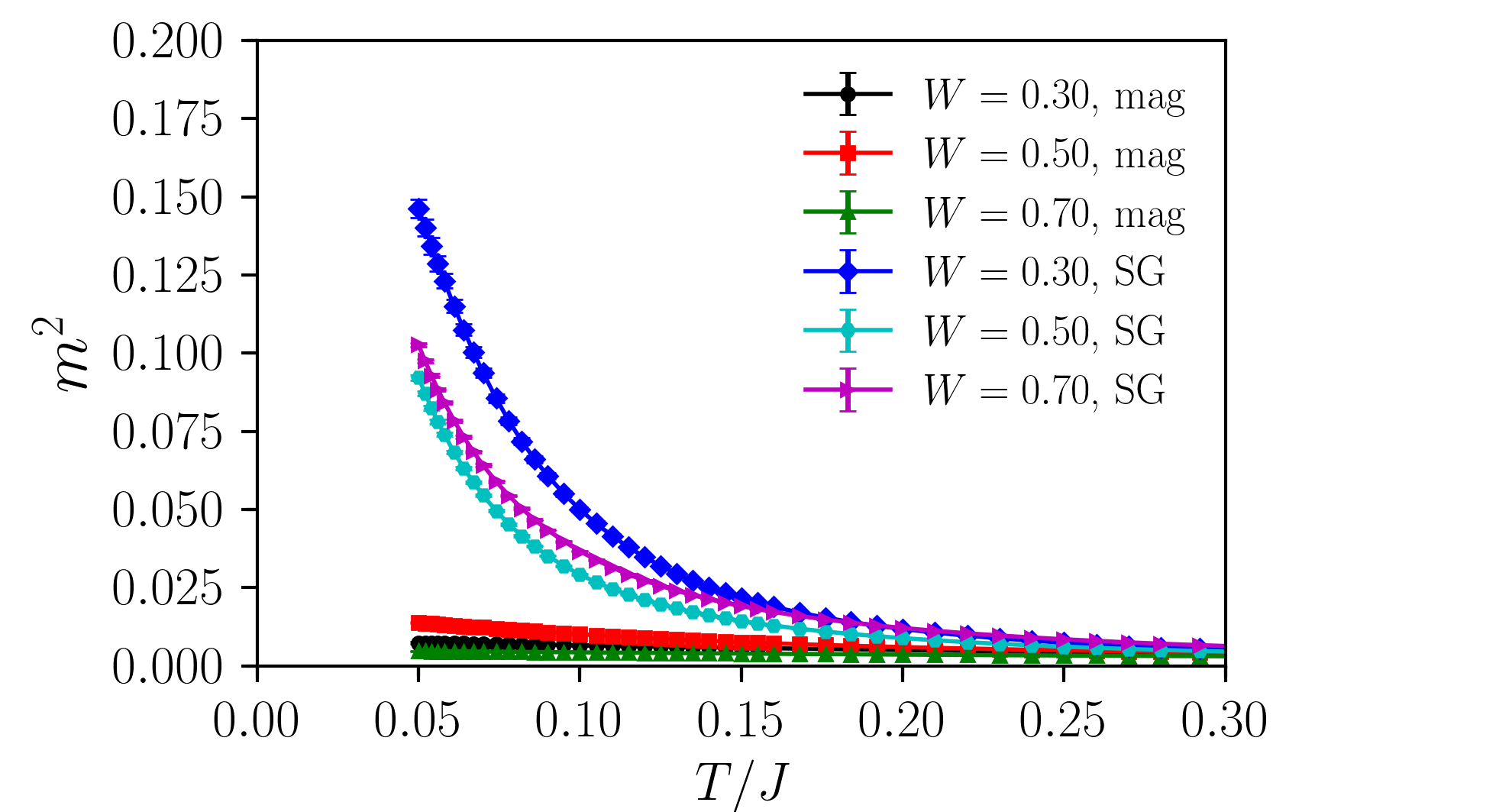}
\includegraphics[width=\columnwidth]{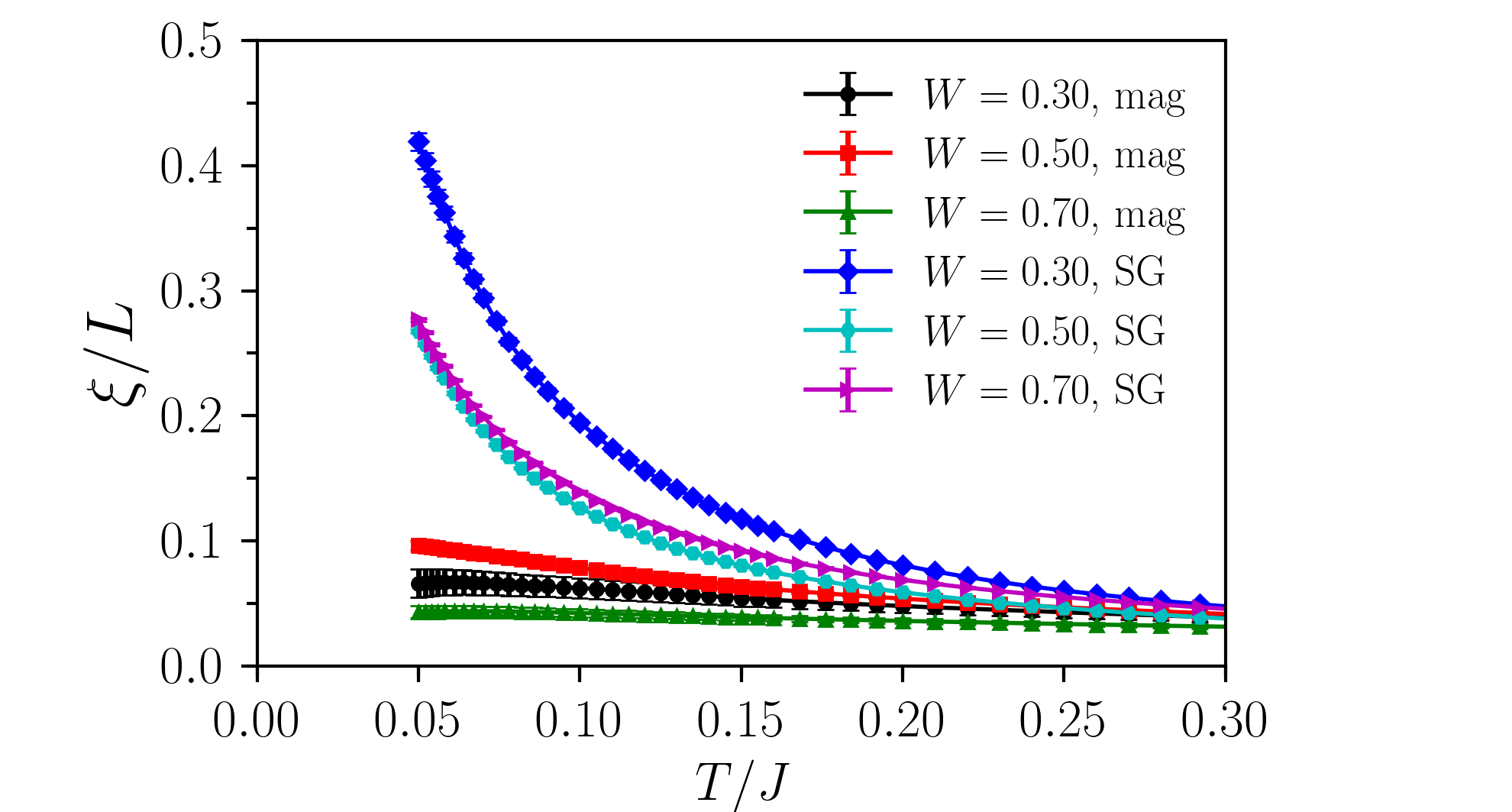}
\caption{
Glassiness: MC results for (a) the order parameters and (b) the correlation lengths for different strength $\Delta$ of Gaussian disorder, comparing magnetic ($120^\circ$) and spin-glass (SG) quantities, at $\alpha=0.05$ and fixed system size $L=36$. The spin-glass quantities are strongly enhanced at low $T$, indicative of the spin-glass ground state.
}
\label{fig:glass}
\end{figure}

Finite-temperature MC results are displayed in Fig.~\ref{fig:glass}, where panel (a) shows both the square of the magnetic order parameter $(m/S)^2$ and the spin-glass order parameter $(m_{\rm SG}/S^2)^2$.
At fixed system size, both the spin-glass order parameter and the spin-glass correlation length $\xi_{\rm sg}$ grow much faster upon cooling than their magnetic counterparts. The behavior of $\xi_{\rm sg}$ is consistent with a divergence as $T\to0$, with $\xi$ remaining finite, as expected for a spin-glass ground state.

\subsection{Non-coplanarity}

The analysis of Sec.~\ref{sec:singresp} shows that the spin configuration remains coplanar in the presence of a single bond defect, consistent with the corresponding numerics. This then also applies to the case of finite bond disorder in the linear-response limit. However, beyond linear response the spin configurations develop non-coplanarities.

\begin{figure}
\includegraphics[width=0.5\columnwidth]{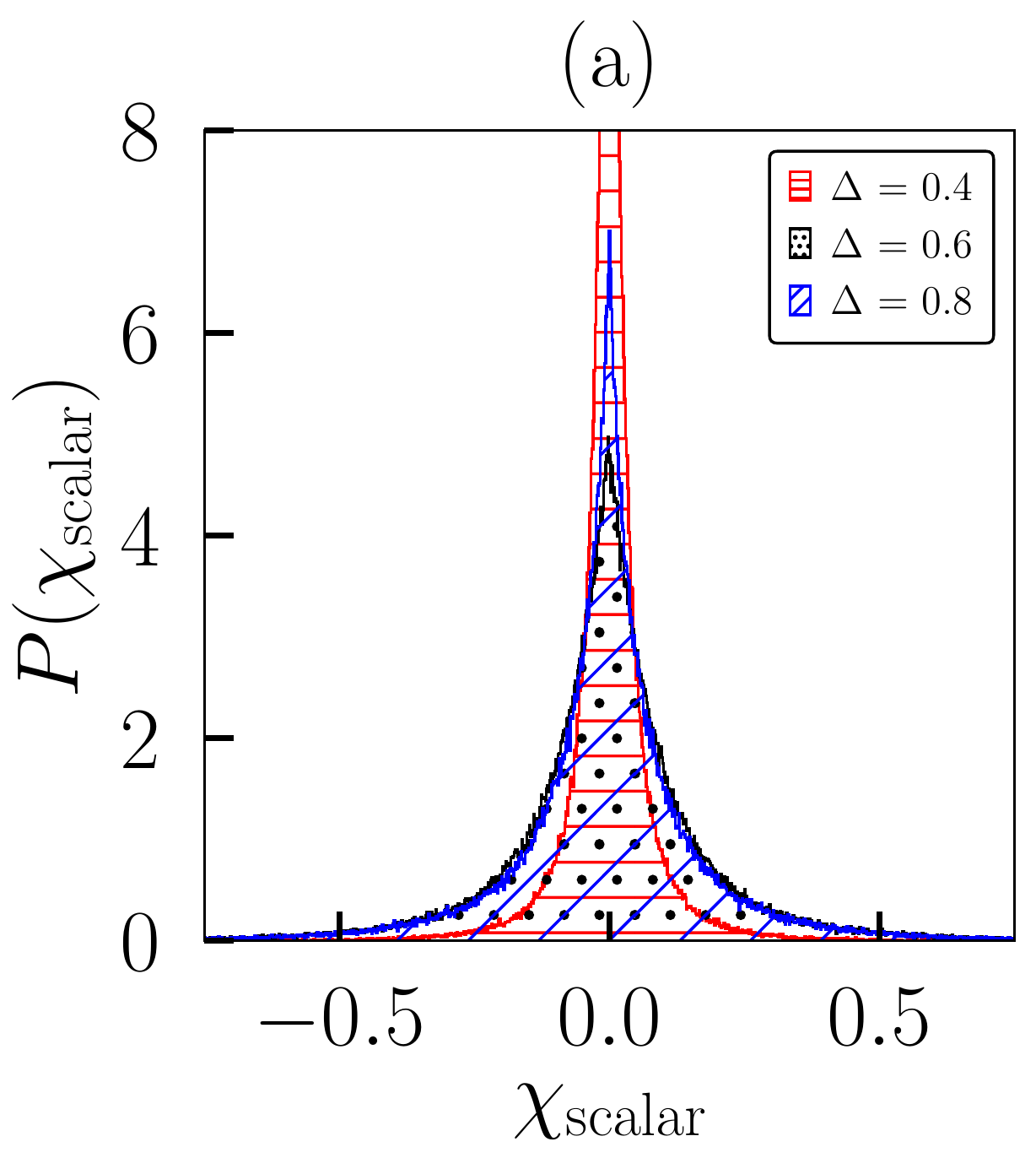}%
\includegraphics[width=0.5\columnwidth]{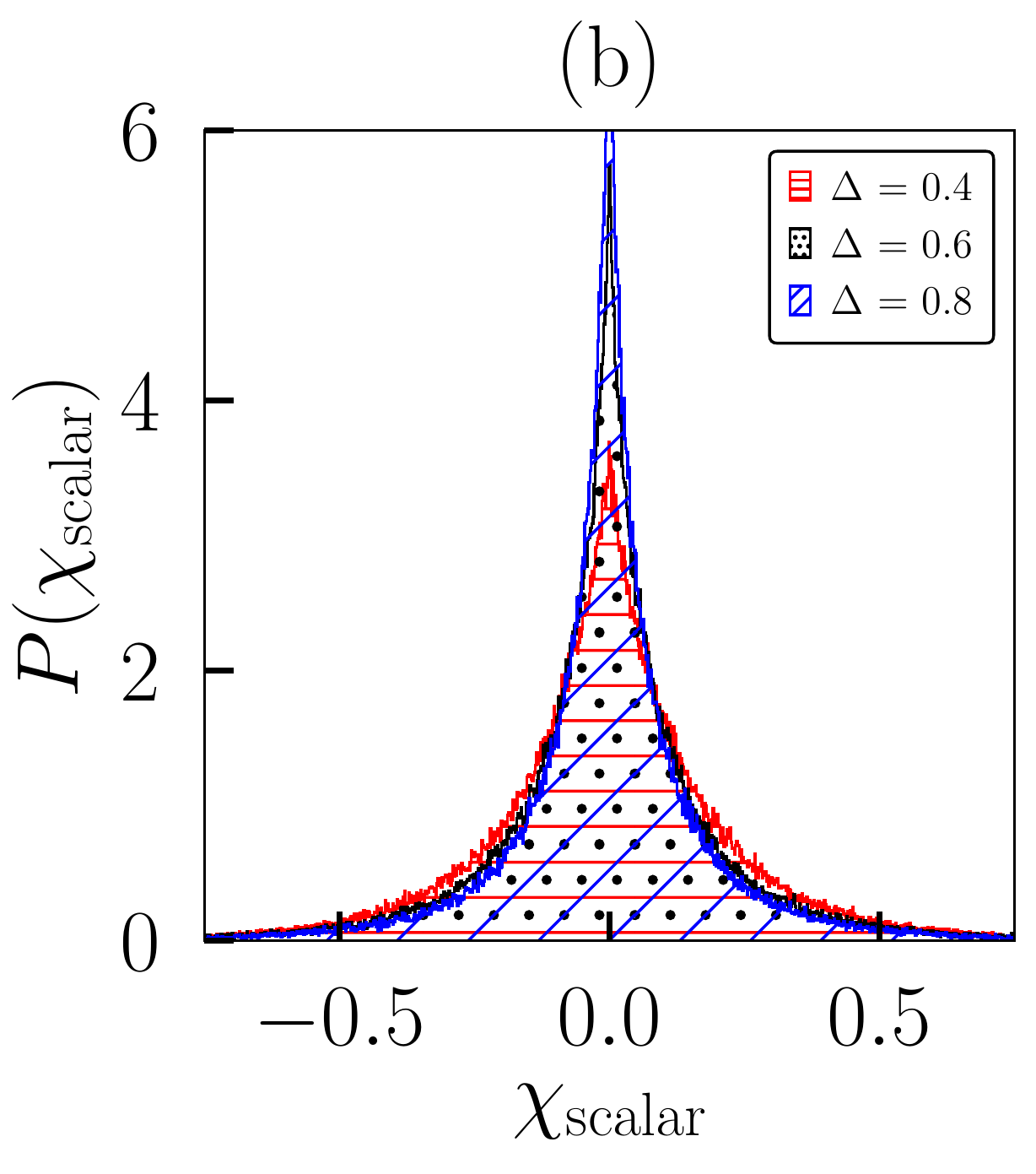}
\vspace*{-8pt}
\caption{Histogram of scalar chiralities $\chi_{ijk}$ for the bond-disordered triangular-lattice Heisenberg model at $T=0$ for different disorder strength $\Delta$ and (a) $\alpha\equiv J_2/J_1=0$ and (b) $\alpha=0.08$.
}
\label{fig:scalarchi}
\end{figure}

We have characterized the degree of non-coplanarity by determining the distribution of scalar chiralities, $\chi_{ijk} = \vec S_i \cdot ( \vec S_j \times \vec S_k )/S^3$ on elementary triangles $ijk$. The distribution is shown in Fig.~\ref{fig:scalarchi}. While the distribution $P(\chi)$ is extremely narrow centered around zero for weak disorder, consistent with a coplanar state in the weak-disorder limit, non-zero chiralities occur for all finite disorder levels $\Delta$. Interestingly, the width of the distribution is non-monotonic as function of $\Delta$: This reflects the fact that, for very strong disorder, exceptionally strong bonds lead to (approximately) collinear spin pairs, rendering the involved triangles coplanar and thus reduding the degree of non-coplanarity compared to intermediate disorder.
